\documentstyle[aps,epsfig,floats,eqsecnum]{revtex}

\def\laq{\raise 0.4ex\hbox{$<$}\kern -0.8em\lower 0.62ex\hbox{$\sim$}}
\def\gaq{\raise 0.4ex\hbox{$>$}\kern -0.7em\lower 0.62ex\hbox{$\sim$}}

\newcommand{\beq}{\begin{equation}}
\newcommand{\eeq}{\end{equation}}
\newcommand{\bea}{\begin{eqnarray}} 
\newcommand{\eea}{\end{eqnarray}}
\newcommand{\ba}{\begin{array}}
\newcommand{\ea}{\end{array}}
\newtheorem{theorem}{Theorem}
\newtheorem{definition}{Definition}

\begin{document}
\draft
\title{\large\bf Signal recycled laser-interferometer
gravitational-wave detectors as optical springs}
\author{Alessandra Buonanno and Yanbei Chen}
\address{Theoretical Astrophysics and Relativity Group,\\
California Institute of Technology,
Pasadena, California 91125, USA}
\vskip 0.2truecm

\maketitle
\begin{abstract}
Using the force-susceptibility formalism of linear quantum
measurements, we study the dynamics of signal recycled 
interferometers, such as LIGO-II. 
We show that, although the antisymmetric mode of motion of the four arm-cavity
mirrors is originally described by a free mass, 
when the signal-recycling mirror is added to the
interferometer, the radiation-pressure force not only
disturbs the motion of that ``free mass'' randomly due to 
quantum fluctuations, but also and more fundamentally,
makes it respond to forces as though
it were connected to a spring with a
specific optical-mechanical rigidity. This oscillatory response
gives rise to a much richer dynamics
than previously known for SR interferometers, which enhances the possibilities
for reshaping the noise curves and, if thermal noise can be pushed low enough,
enables the standard quantum limit to be beaten. 
We also show the possibility of using servo systems to suppress the
instability associated with the optical-mechanical interaction without
compromising the sensitivity of the interferometer. 
\end{abstract}
\vskip 0.2truecm
\centerline{\small  PACS No.: 04.80.Nn, 95.55.Ym, 42.50.Dv, 03.65.Bz.
\hspace{0.5 cm} GRP/00/554}

\section{Introduction}
\label{sec1}

\begin{figure}[ht]
\vskip -0.4truecm
\begin{center}
\epsfig{file=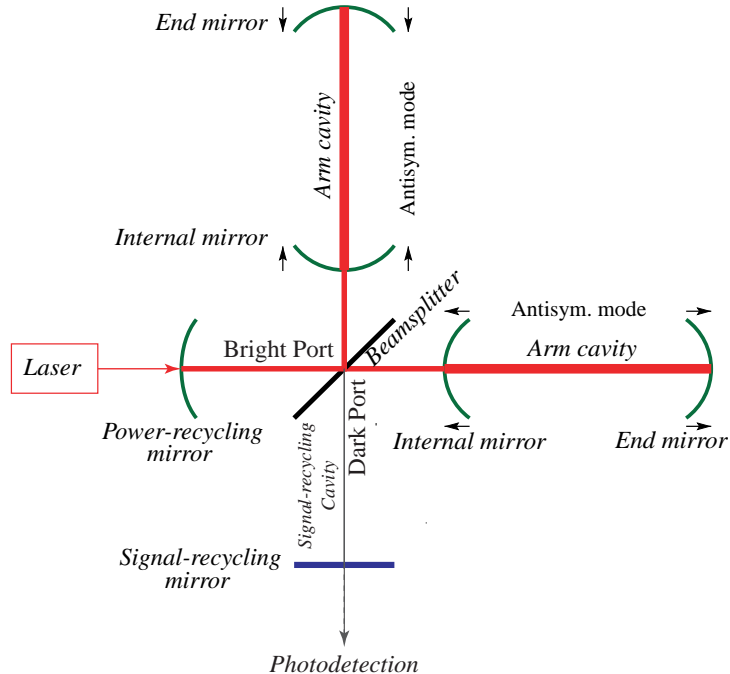,width=0.6\textwidth,angle=0}
\vskip 0.2truecm
\caption{\sl Schematic diagram of a signal recycled
interferometer such as LIGO-II. The antisymmetric mode of motion of the four
arm-cavity mirrors (marked by arrows) is monitored by laser
interferometry. A signal-recycling mirror is used to feed the signal
light back into the arm cavities, while a power-recycling mirror is introduced  
to feed back into the arm cavities the unused laser light coming out 
the bright port.}
\label{Fig1}
\end{center}
\end{figure}

Next year a network of broadband ground-based laser interferometers,
aimed to detect gravitational waves (GWs) in the frequency band
$10-10^4\,$Hz, will begin operations.
This network is composed of the Laser Interferometer Gravitational-wave Observatory
(LIGO), VIRGO (whose operation will begin in 2004), 
GEO\,600, and TAMA\,300 \cite{Inter}.
Given the anticipated noise spectra
and the current estimates of gravitational waves from
various astrophysical sources \cite{Sour},
it is plausible but not probable that gravitational waves
will be detected with the first generation of interferometers.
The original conception of LIGO included an upgrade
of LIGO to sensitivities at which it is probable
to detect a rich variety of gravitational waves \cite{Sour}.
The LIGO Scientific Collaboration (LSC) \cite{GSSW99} is currently planning
this upgrade to begin in  2006. This second stage includes: (i)
improvement of the seismic isolation system to push
the seismic wall downward in frequency
to 10\,Hz, (ii) improvement of the suspension system
to lower the noise in the band between $\sim \! 10$\,Hz and
$\sim \! 200$\,Hz, (iii) increase (decrease) of light
power (shot noise) circulating in the
arm cavities ($ \sim \! 1$\,MWatt), (iv) improvement
in the optics so that they can handle the increased
laser power, and (v) introduction of an extra mirror,
called a signal-recycling (SR) mirror, at the dark-port output.
This upgraded configuration of LIGO (``advanced interferometer'')
is sometimes called LIGO-II and its design is sketched in Fig.~\ref{Fig1}.

The SR mirror (see Fig. \ref{Fig1}) sends the signal coming out
the dark port back into the arm cavities; in this
sense it \emph{recycles} the \emph{signal}.
\footnote{~The configuration of LIGO-II will also include a power-recycling (PR) mirror  
between the laser and the beamsplitter (see Fig.~\ref{Fig1}). This mirror {\it recycles}
back into the arm cavities the unused laser light coming out 
the bright port and increases the light power at the beamsplitter.
Besides this effect, the presence of the PR mirror does not affect the 
derivation of the quantum noise at the dark-port output. 
Therefore, although in our analysis we assume high light power, 
we do not need to take into account the PR mirror 
in deducing the interferometer's input-output relation.} 
The optical system composed of the SR cavity and the arm cavities forms a composite
resonant cavity, whose eigenfrequencies and quality factors can be controlled by
the position and reflectivity of the SR mirror.
Near its eigenfrequencies (resonances), 
the device can gain sensitivity. In fact, the initial
motivation for introducing the SR cavity was based on
the idea of using this feature to reshape the
noise curves, enabling the interferometer to work
either in broadband or in narrowband
configurations, and improving in this way
the observation of specific GW astrophysical
sources \cite{Sour}.
Historically, the first idea for a narrowband configuration,
so-called {\it synchronous} or {\it resonant recycling},
was due to Drever \cite{D82} and was subsequently analyzed by Vinet et
al.\ \cite{VMMB88}. It used a different optical 
topology from Fig.~\ref{Fig1}. 
The original idea for the optical topology of Fig.~\ref{Fig1} 
was due to Meers \cite{M88}, who proposed its use for
{\it dual recycling} -- a scheme 
which by recycling the signal light {\it increases}
the storage time of the signal inside the interferometer and lowers the shot noise.
Later, Mizuno et al.\ \cite{MSNCSRWD93,M95,H99} proposed another scheme called Resonant
Sideband Extraction (RSE), which also uses the optical topology of 
Fig.~\ref{Fig1} but adjusts the SR mirror so that 
the storage time of the signal inside the interferometer {\it decreases} 
while the observation bandwidth {\it increases}. 
In general, by choosing appropriate detunings 
\footnote{~By detuning of the SR cavity we mean the phase gained by the
carrier frequency in the SR cavity, see Sec.~\ref{sec3.1} for details.}
of the SR cavity, the optical configuration can be in either of the two regimes,
or in between. These schemes have been experimentally
tested by Freise et al.\ \cite{FHSMSLWSRWD00} with the 30\,m 
laser interferometer in Garching (Germany), and by Mason
\cite{M01} on a table-top experiment at Caltech (USA).

All the above mentioned theoretical analyses and experiments
of SR interferometers \cite{D82,VMMB88,M88,MSNCSRWD93,M95,H99,FHSMSLWSRWD00,M01}
refer to configurations with low laser power, 
for which the radiation pressure on the
arm-cavity mirrors is negligible and the noise spectra
are dominated by shot noise.
However, when the laser power is increased, the shot noise decreases
while the effect of radiation-pressure fluctuation increases.
LIGO-II has been planned to work at a laser power for which the
two effects are comparable in the observation band $10$--$200$\,Hz \cite{GSSW99}. 
Therefore, to correctly describe the quantum optical noise in LIGO-II, 
the results so far obtained in the literature 
\cite{D82,VMMB88,M88,MSNCSRWD93,M95,H99,FHSMSLWSRWD00,M01} 
must be complemented by a thorough investigation of the influence of the
radiation-pressure force on the mirror motion.

Until recently the LIGO-II noise curves were computed using a
semiclassical approach \cite{GSSW99}, which, although capable
of estimating the shot noise, is unable to
take into account correctly the effects of radiation-pressure fluctuations. 
Very recently, building on earlier work of Kimble, Levin, Matsko,
Thorne and Vyatchanin (KLMTV for short) \cite{KLMTV00}, which describes 
the initial optical configuration of LIGO/TAMA/VIRGO interferometers
(so-called conventional interferometers) within a full
quantum-mechanical approach, we investigated the SR optical configuration 
(Fig.~\ref{Fig1}) \cite{BC1,BC2}. Our analysis revealed important new
properties of SR interferometers, including:
(i) the presence of correlations between shot noise
and radiation-pressure noise, (ii) the possibility of beating the
standard quantum limit (SQL) by a modest amount,
roughly a factor of two over a bandwidth of $\Delta f\!\sim\! f$
\footnote{~This performance refers only to the quantum optical noise. 
The total noise, 
which includes also all the other sources of noise, 
such as seismic and thermal noise, can
beat the SQL only if thermoelastic noise \cite{BGV00} can
also be pushed below the SQL.} 
and (iii) the presence of instabilities in the
optical-mechanical system formed by the optical fields
and the arm-cavity mirrors.
We also noticed \cite{BC2} that the way the SQL is beaten in 
SR interferometer is quite different from standard quantum-nondemolition (QND)
techniques \cite{B68-70s} based on building up 
correlations between shot noise and radiation-pressure noise by (i)
injecting squeezed vacuum into an interferometer's dark port
\cite{SFD} and/or (ii) introducing two kilometer-long filter cavities
into the interferometer's output port~\cite{HFD,KLMTV00} and applying
homodyne detection on the filtered light. Indeed, our analyses suggest
that the improvement in the noise curves comes largely from the resonant
features introduced by the SR cavity: whereas the amplitude of the classical
output signal is amplified near the resonances,
the output quantum fluctuation is not strongly affected by them.
This way of using resonances to beat the SQL was first proposed by
Braginsky, Khalili and colleagues in their scheme of ``optical bar'' GW
detectors \cite{OB}, where similarly the test mass is effectively an
oscillator whose restoring force is provided by in-cavity optical fields.
For an ``optical bar'' the free-mass SQL is irrelevant 
and we can beat the free-mass SQL using classical
techniques of position monitoring \cite{OB}.

In Ref.~\cite{BC2} our analysis was mainly focused on
determining the input--output relations for
the electromagnetic quadrature fields in a SR interferometer,
and evaluating the corresponding noise spectral density.
The resonant features of the whole device
were discussed only briefly. In the present paper we
give a detailed description of the dynamics
of the system formed by the optical fields
and the mirrors, we discuss the origin of
the resonances and their possible instabilities,
and we analyze the suppression of the instabilities
by an appropriate control system.
In our analysis we have found the Braginsky-Khalili 
formalism for linear quantum measurements
\cite{BK92} very powerful and intuitive, and
we use it throughout this paper.

This paper is divided into two parts: the formalism and its
application. In Sec.~\ref{sec2} we
introduce the force-susceptibility formalism and discuss some general
features of linear quantum-measurement devices. In particular,  
after briefly commenting in Sec.~\ref{subsec2.1} 
on general quantum-measurement systems, we derive in Sec.~\ref{subsec2.2}
the equations of motion for {\it linear} quantum-measurement devices; 
in Sec.~\ref{subsec2.3} we write down a set of conditions on the susceptibilities of 
linear quantum-measurement systems; in Sec.~\ref{subsec3.2} we use these conditions to
construct an effective description of a quantum-measurement process
which allows us to identify in a straightforward way the shot noise and
the radiation-pressure noise. 
In the subsequent sections we apply the formalism developed in
Sec.~\ref{sec2} to SR interferometers. In 
Sec.~\ref{sec3} we show that SR interferometers can 
be described by the force-susceptibility formalism and we derive their 
equations of motion,
pointing out the existence of a ``ponderomotive rigidity''.
In Sec.~\ref{sec4} we
discuss in detail the oscillatory behavior of the system induced by the
ponderomotive rigidity, its resonances and
instabilities. In Sec.~\ref{sec5} we describe the 
suppression of the instability by a feed-back control system which 
does not compromise the sensitivity. Finally, Sec.~\ref{sec6} summarizes 
our main conclusions. As a foundation for our linear analysis of SR interferometers
we summarize in Appendix~\ref{appA} some general
properties of linear quantum-mechanical systems.

\section{Quantum-measurement systems}
\label{sec2}

\subsection{General conditions defining a measurement system}
\label{subsec2.1}

\begin{figure}[ht]
\vskip -0.5truecm
\begin{center}
\epsfig{file=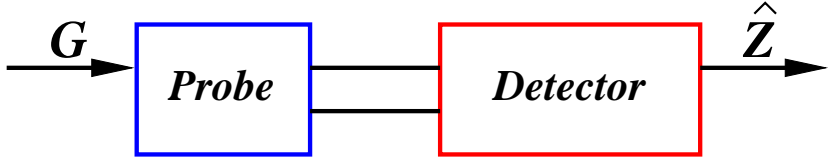,width=0.13\textwidth,angle=-90}
\vskip 0.2truecm
\caption{\sl Schematic diagram of a measurement device. $G$ is the
classical observable acting on the probe that we want to measure, 
and $\widehat{Z}$ is the detector's observable which describes
the output of the measurement system.}
\label{Fig0}
\end{center}
\end{figure}

Following Braginsky and Khalili \cite{BK92}, we define a
{\it measurement process} as a transformation from some original classical
observable which is {\it unknown}, e.g., the gravitational-wave
amplitude, into another classical observable which is {\it known},
e.g., the data stored in the computer. Generally, the system which
implements this process is composed of a probe ${\cal P}$,
which is directly coupled to the classical observable to be measured
(for interferometers this is the antisymmetric
mode of motion of the four arm-cavity mirrors, see Sec.~\ref{identify}),
and the detector ${\cal D}$, which couples to the probe 
and produces the output observable
(for interferometers this is the optical system
and the photodetector). A measurement system is drawn schematically in Fig.~\ref{Fig0}.
Because the probe and the detector are quantum mechanical systems,
the overall device is called a {\it quantum-measurement device}.
The output observable $\widehat{Z} = {\cal S} + \widehat{\cal Q}$ 
contains a classical part ${\cal S}$, which depends on the classical observable
$G$ to be measured, and some quantum noise $\widehat{\cal Q}$
due to the probe, the detector and their mutual interaction.

According to the statistical interpretation of Quantum Mechanics
\cite{vonneumann}, the output of a quantum-measurement process at
different times should be {\it simultaneously measurable}. One
sufficient condition for {\it simultaneous measurability} is that the
Heisenberg operators of the output observable, $\widehat{Z}(t)$,
satisfy \footnote{
~We refer to this condition as sufficient since for observables
that do not satisfy this condition,
there may still exist a subspace of the Hilbert space of the system in
which these observables are simultaneously measurable. }

\beq
\left[\widehat{Z}(t_1),\widehat{Z}(t_2)\right]=0\quad\forall t_1,\,t_2\,.
\label{commute}
\eeq
Henceforth, we shall regard Eq.~(\ref{commute}) as the {\it condition of
simultaneous measurability}. 
Although the condition (\ref{commute}) was originally 
introduced by Braginsky et al.\ \cite{B68-70s,BK92} 
as the definition of quantum-nondemolition (QND) observables (see also 
Refs.~\cite{unruh,CTDSZ80,qnd}), we introduce and use it for different purposes, 
as will become clear in the following.
If the condition (\ref{commute}) is satisfied, 
then any sample of data
$\left\{\widehat{Z}(t_1),\widehat{Z}(t_2),\ldots,\widehat{Z}(t_n)\right\}$ 
can be stored directly as bits of classical data in a classical storage medium,
and any noise from subsequent processing of the signal can be made arbitrarily 
small, i.e.\ {\it all quantum noises} are included in the quantum fluctuations of
$\widehat{Z}(t)$.
We want to discuss the {\it simultaneous measurability} condition
(\ref{commute}) more deeply by pointing out the following relation,
which was also in part discussed by Unruh \cite{unruh} and 
Caves, Thorne, Drever, Sandberg and
Zimmermann in Sec.~IV of Ref.~\cite{CTDSZ80}, and reviewed 
subsequently in Ref.~\cite{qnd}, although from a different point of view.

\vspace{0.2cm}
\noindent 
{\bf Simultaneous-Measurability -- Zero-Response Relation:}
\emph{For a Quantum Measurement Device (QMD), the simultaneous measurability
condition for the output $\widehat{Z}(t)$, i.e.\
$[\widehat{Z}(t_1),\widehat{Z}(t_2)]=0\quad\forall
t_1,\,t_2$, is equivalent to requiring that if the device is coupled to an
external system via an interaction Hamiltonian of the form
$V(\widehat{Z},\widehat{\cal E})$ where $V$ is an arbitrary function and 
$\widehat{\cal E}$ belongs to the external system,
then the back action on the QMD does \underline{not} alter
the evolution of the output observable $\widehat{Z}$.
}

\vspace{0.2cm}
\emph{Proof of necessity.}
\footnote{~A similar calculation was carried out by Caves
et al.\ in Sec.~IV of Ref.~\cite{CTDSZ80}.}
Let us suppose that our QMD with output $\widehat{Z}$
evolves under a Hamiltonian $\widehat{H}_{\rm QMD}$, and that 
$[\widehat{Z}(t),\widehat{Z}(t')]=0$ for all $t,t'$.
Now let us
couple it to an arbitrary external system with  Hamiltonian 
$\widehat{H}_{\rm EXT}$ via a
generic interaction term $V(\widehat{Z},\widehat{\cal E})$ as specified above,
where $\widehat{\cal E}$ is an observable of the
external system. The total Hamiltonian is
\beq
\widehat{H}=
\left(\widehat{H}_{\rm QMD}+\widehat{H}_{\rm EXT}\right) +
V(\widehat{Z},\widehat{\cal E})\,.
\eeq
If we treat the two terms in the bracket as the zeroth-order
Hamiltonian and the interaction Hamiltonian
$V(\widehat{Z},\widehat{\cal E})$ as a perturbation,
by applying the results derived in the Appendix
[see Eq.~(\ref{a15})] we can write the Heisenberg operator of the output variable
$\widehat{Z}$ as,
\bea
&& \widehat{Z}_{\rm pert}(t)=\widehat{Z}(t) +
\frac{i}{\hbar}\int_{-\infty}^t dt_1 \left[V(\widehat{Z}(t_1),
\widehat{\cal E}(t_1)),
\widehat{Z}(t)\right]+ \nonumber \\
&& \left(\frac{i}{\hbar}\right)^2 \int_{-\infty}^t dt_1 \int_{-\infty}^{t_1} dt_2
\left[V(\widehat{Z}(t_2),\widehat{\cal E}(t_2)),
\left[V(\widehat{Z}(t_1),\widehat{\cal E}(t_1)),
\widehat{Z}(t)\right]\right] + \cdots \,,
\label{perturbed}
\eea
with higher order terms of the form 
[see Eq.~(\ref{a15})]:
\beq
\left[V(\widehat{Z}(t_n),\widehat{\cal E}(t_n)),
\left[\cdots,\left[V(\widehat{Z}(t_2),\widehat{\cal E}(t_2)),
\left[V(\widehat{Z}(t_1),\widehat{\cal E}(t_1)),
\widehat{Z}(t)\right]\right]\cdots\right]\right]\,.
\label{ht}
\eeq
Here $\widehat{Z}(t)$ and $\widehat{\cal E}(t)$
evolve under the Hamiltonians
$\widehat{H}_{\rm QMD}$ and $\widehat{H}_{\rm EXT}$, respectively.
Because they belong to two different Hilbert spaces we have 
$[\widehat{Z}(t), \widehat{\cal E}(t')]=0$ for all $t,t'$. 
By assumption, we also have
$[\widehat{Z}(t_1),\widehat{Z}(t_2)]=0\quad\forall t_1,\,t_2$.
Using these two facts, we obtain 
$[V(\widehat{Z}(t_1),\widehat{\cal E}(t_1)),\widehat{Z}(t_2)]=0
\quad \forall t_1,\,t_2$, and then using Eq.~(\ref{perturbed}) 
we derive $\widehat{Z}_{\rm pert}(t)=\widehat{Z}(t)$.
This means that the evolution of $\widehat{Z}$ is not affected by the
kind of external coupling we introduced.

\emph{Proof of sufficiency.}
Let us suppose the evolution of $\widehat{Z}$ is not affected by
\underline{any} external system of the form specified above.
Then, in particular, it must be true for the simple interaction Hamiltonian
$V(\widehat{Z},\widehat{\cal E}) = -\alpha \widehat{Z}\,{\cal E}$,
where  $\alpha$ is some coupling constant which can vary continuously, e.g.,  
in the interval $(0,1]$, and we choose a classical external coupling ${\cal E}$. 
In this particular case Eq.~(\ref{perturbed}) becomes

\beq
\widehat{Z}_{\rm pert}(t)=
\widehat{Z}(t)
-\alpha\, \frac{i}{\hbar}\,\int_{-\infty}^t dt_1
\left[\widehat{Z}(t_1),\widehat{Z}(t)\right]\,
{\cal E}(t_1) + O(\alpha^2)\,,
\label{alpha}
\eeq
with higher order terms of the form:
$\alpha^n\,[\widehat{Z}(t_n),[ \cdots, [\widehat{Z}(t_2),
[\widehat{Z}(t_1),\widehat{Z}(t)]] \cdots]]$.
By assumption the LHS of Eq.~(\ref{alpha}) does not change when we vary
$\alpha$. The RHS of Eq.~(\ref{alpha}) is a power series in $\alpha$, and using the uniqueness
of the Taylor expansion, we deduce that all the terms beyond the zeroth order 
should vanish separately. In particular, the first-order term should vanish
and we conclude that $[\widehat{Z}(t),\widehat{Z}(t')]=0$ for all $t,t'$. $\Box$

Let us comment on two interesting aspects of the 
{Simultaneous-Measurability -- Zero-Response Relation} given above.
\begin{itemize}
\item This relation links the abstract quantum mechanical idea of
{\it simultaneous measurability} to the classical dynamics of the measurement
device, yielding a simple criterion for the quantum-classical 
transition: the observable which corresponds to the classical output
variable should have no
response to external perturbations directly coupled
to it.
\footnote{~By directly coupling to $\widehat{Z}$ we mean the
interaction Hamiltonian is of the form
$V(\widehat{Z},\widehat{\cal E})$,
since only this form guarantees that $\widehat{Z}$ is the only 
observable of the device that influences the interaction.}
We shall use this criterion in our
analysis of linear systems in Sec.~\ref{subsec2.3}.
\item
This relation is also interesting conceptually.
In practice, the result of every measurement is read out by coupling 
the measurement device to another system,
and the boundary between the ``measurement'' (still part of the QMD)
and the ``data analysis'' (external to the QMD) occurs 
at a ``stage'' at which no possible direct coupling to the output observable
could change the evolution of the output observable itself. Otherwise at that 
stage the ``external coupling'' should still be
considered as part of the measurement device.
\end{itemize}

Before ending this section, let us compare the point of view followed 
in this section to the one pursued in previous QND analyses 
\cite{unruh,CTDSZ80,qnd}, especially 
Sec.~IV of Ref.~\cite{CTDSZ80}. 
The authors of Refs.~\cite{CTDSZ80,qnd} followed two steps 
in their discussion. First, they searched for a class of observables $\widehat{A}(t)$ 
of a quantum-mechanical system that can be monitored
without adding fundamental noise, deducing a condition 
for  $\widehat{A}(t)$ that coincides with
Eq.~(\ref{commute}). They called such
observables QND observables. 
Secondly, they found appropriate interaction Hamiltonians 
describing the coupling between $\widehat{A}(t)$ and a 
measuring apparatus that do not disturb
the evolution of  $\widehat{A}(t)$  during the measurement process. 
However, in Refs.~\cite{CTDSZ80,qnd} there is no 
clear distinction between what we call the detector and 
the external measurement system; these two systems are referred 
to together as the measuring apparatus. Thus, the observable 
$\widehat{A}(t)$ does not necessarily coincide with the output 
$\widehat{Z}(t)$ of our probe-detector system, and for this reason   
we prefer not to call it a QND observable in the sense 
of Refs.~\cite{unruh,CTDSZ80,qnd}.

As a final remark, we note that whereas in Refs.~\cite{CTDSZ80,qnd} 
the measuring apparatus and the interaction Hamiltonian are 
indispensable parts of a measurement process, 
in this paper, by distinguishing the detector from the  
external  system, we use the latter only as part of a {\it gedanken}
experiment, by which we clarify the relation between 
simultaneous measurability and the response to external couplings, which
will lead to useful properties of linear quantum-measurement devices in 
Sec.~\ref{subsec2.3}.

\subsection{Equations of motion of a linear quantum-measurement system:
The force-susceptibility formalism}
\label{subsec2.2}

Starting in this section we shall focus on linear measurement systems.
We shall see in Sec.~\ref{sec3} that GW 
interferometers belong to this class of devices. Our analysis has been inspired
by the formalism of linear quantum-measurement theory introduced
by Braginsky and Khalili (Chaps. V, VI and VII of Ref.~\cite{BK92}) and is based on the
force-susceptibility description of linearly coupled systems under
linearly applied classical forces (see, e.g., Sec.~6.4 of Ref.~\cite{BK92}).

\begin{figure}
\vskip -0.5truecm
\begin{center}
\epsfig{file=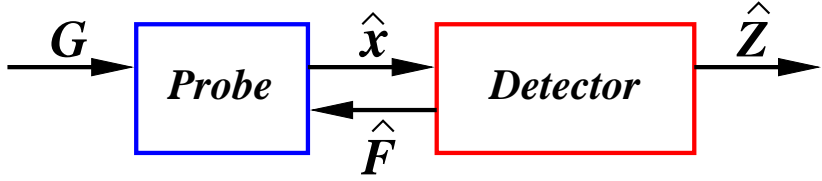,width=0.15\textwidth,angle=-90}
\vskip 0.2truecm
\caption{\sl Schematic diagram of a linear measurement
system. $G$ is the external
classical force acting on the probe that we want to measure,
$\widehat{x}$ is the linear observable of the probe,
$\widehat{F}$ is the linear observable of the detector
which describes the back-action force on the probe, and $\widehat{Z}$ is
the linear observable of the detector which describes the output of the
overall measurement system.}
\label{Fig2}
\end{center}
\end{figure}

In a {\it linear measurement process}, the device acts
linearly and is linearly coupled to the classical observable to be 
measured (see the Appendix for a precise definition of linear systems).
We suppose that the device can be
artificially divided into two linearly coupled, but otherwise
independent subsystems: the probe, which is subject to the
external classical force we want to measure,
and the detector, which yields a classical output.
More specifically, in our Hamiltonian system the probe is coupled
to the external classical force $G$ by $-\widehat{y}\,G$,
where $\widehat{y}$ is some linear observable of the probe,
while the probe and the detector are coupled by a term $-\widehat{x}\,\widehat{F}$,
where $\widehat{x}$ is a generalized (linear) displacement of the probe,
and $\widehat{F}$ is a linear observable of the detector which describes its
back-action force on the probe.
In general, the observable $\widehat{x}$ to which the external force is
coupled and the observable $\widehat{y}$ 
that the detector directly measures might not be the same.
However, in our idealized model of GW interferometers (Sec.~III below),
$\widehat{x}$ and $\widehat{y}$ are actually
the same observable, namely the generalized coordinate of the antisymmetric
mode of motion of the four arm-cavity mirrors (see Fig.~\ref{Fig1} and
Sec.~\ref{identify}), and $\widehat{F}$ is the
radiation-pressure force acting on this mode. 
Henceforth, we shall impose $\widehat{y} \equiv \widehat{x}$.
Finally, we denote by $\widehat{Z}$ the linear observable of the detector
which describes the output of the entire device. A sketchy representation
of the measurement device is drawn in Fig.~\ref{Fig2}.
The linear observables $\widehat{x}$ 
describing the probe ${\cal P}$ and 
$\widehat{Z}$, $\widehat{F}$ describing the detector ${\cal D}$
belong to two different Hilbert spaces $\cal{H}_{\cal P}$ and
$\cal{H}_{\cal D}$, respectively, 
and the Hilbert space of the combined system is 
$\cal{H}_{\cal P} \otimes \cal{H}_{\cal D}$.
The Hamiltonian is given by
\beq
\label{2.1}
\widehat{H}
= \left[\left(\widehat{H}_{\cal P}-\widehat{x}\,G\right) + \widehat{H}_{\cal D} \right]
- \widehat{x}\,\widehat{F}\,.
\eeq

We shall now derive the equations of motion of the system composed 
of the linear observables $\widehat{x}$, $\widehat{Z}$ and $\widehat{F}$.
As a first step in our calculation, we regard the Hamiltonians $\widehat{H}_{\cal P} -
\widehat{x}\,G$ and $\widehat{H}_{\cal D}$  as zeroth order Hamiltonians
for the subsystems ${\cal P}$ and ${\cal D}$, respectively, and we treat
$- \widehat{x}\,\widehat{F}$ as a linear coupling between ${\cal P}$ and ${\cal D}$.
Working in the Heisenberg picture, we obtain the following equations
[see Theorem 4 of the Appendix
and Eqs.~(\ref{a20}), (\ref{a21})]:
\bea
\label{z1}
\widehat{Z}^{(1)}(t)&=&\widehat{Z}^{(0)}(t)+
\frac{i}{\hbar}\int_{-\infty}^{t}dt'\,C_{Z^{(0)}F^{(0)}}(t,t')\,\widehat{x}^{(1)}(t')\,,
\\
\label{f1}
\widehat{F}^{(1)}(t)&=&\widehat{F}^{(0)}(t)+
\frac{i}{\hbar}\int_{-\infty}^{t}dt'\,C_{F^{(0)}F^{(0)}}(t,t')\,\widehat{x}^{(1)}(t')\,,
\\
\label{x1}
\widehat{x}^{(1)}(t)&=&\widehat{x}^{(G)}(t)+
\frac{i}{\hbar}\int_{-\infty}^{t}dt'\,C_{x^{(G)}x^{(G)}}(t,t')\,\widehat{F}^{(1)}(t')\,.
\label{2.2} 
\eea
Here $C_{AB}(t,t')$ is a complex number (C-number), called the (time-domain)
susceptibility, and is defined by Eq.~(\ref{a18}) of the Appendix, i.e.\
\beq
C_{AB}(t,t')\equiv\left[\widehat{A}(t),\,\widehat{B}(t')\right]\,.
\label{susc}
\eeq
[Henceforth, we shall often use the expressions \emph{different-time commutator}
and \emph{time-domain susceptibility} interchangeably.]
The superscript ${(1)}$ in Eqs.~(\ref{z1})--(\ref{x1}) denotes time evolution under the total Hamiltonian
$\widehat{H}$ [Eq.~(\ref{2.1})], the superscript ${(0)}$ on $\widehat{F}(t)$ and
$\widehat{Z}(t)$ denotes time evolution under the free Hamiltonian
of the detector $\widehat{H}_{\cal D}$, while the superscript ${(G)}$ on
$\widehat{x}(t)$ refers to the time evolution under the
Hamiltonian $\widehat{H}_{\cal P}-\widehat{x}\,G$, which describes
the probe under the sole influence of $G(t)$.

As a second step, we want to relate $\widehat{x}^{(G)}(t)$
to $\widehat{x}^{(0)}(t)$, which evolves under the free probe Hamiltonian
$\widehat{H}_{\cal P}$. Using Theorem 3 in the Appendix and Eqs.~(\ref{a17}), (\ref{a18}),
we deduce
\beq
\widehat{x}^{(G)}(t)=\widehat{x}^{(0)}(t)+
\frac{i}{\hbar}\int_{-\infty}^{t}dt'\,C_{x^{(0)} x^{(0)}}(t,t')\,G(t')\,.
\label{2.3}
\eeq
Noticing from Eq.~(\ref{2.3}) that
$\widehat{x}^{(G)}$ differs from $\widehat{x}^{(0)}$
by a time dependent C-number, we get
$C_{x^{(G)}x^{(G)}}(t,t')=C_{x^{(0)}x^{(0)}}(t,t')$. Using this fact and inserting Eq.~(\ref{2.3}) into
Eq.~(\ref{2.2}), we can relate the Heisenberg operators evolving under
the full Hamiltonian $\widehat{H}$ to
those evolving under the free Hamiltonians of the probe and the detector
$\widehat{H}_{\cal P}$ and $\widehat{H}_{\cal D}$ separately:
\bea
\label{2.4}
\widehat{Z}^{(1)}(t)&=&\widehat{Z}^{(0)}(t)+
\frac{i}{\hbar}\int_{-\infty}^{t}dt'\,C_{Z^{(0)} F^{(0)}}(t,t')\,\widehat{x}^{(1)}(t')\,, \\
\label{2.5}
\widehat{F}^{(1)}(t)&=&\widehat{F}^{(0)}(t)+
\frac{i}{\hbar}\int_{-\infty}^{t}dt'\,C_{F^{(0)} F^{(0)}}(t,t')\,\widehat{x}^{(1)}(t')\,, \\
\label{2.6}
\widehat{x}^{(1)}(t)&=&\widehat{x}^{(0)}(t)
+\frac{i}{\hbar}\int_{-\infty}^{t}dt'\,C_{x^{(0)} x^{(0)}}(t,t')\,
[G(t')+\widehat{F}^{(1)}(t')]\,.
\eea
A quantity of special interest for us is the displacement induced on a free probe
(without any influence of the detector) by $G(t)$, namely the second term
on the RHS of Eq.~(\ref{2.3}). 
For a GW interferometer this displacement is $L\,h(t)$, where $L$ is the arm-cavity length and 
$h(t)$ is the differential strain induced by the gravitational wave on the free arm-cavity mirrors 
(the difference in strain between the two arms).
In our notation we denote this quantity by 
\beq
L\,h(t) = \frac{i}{\hbar}\int_{-\infty}^{t}dt'\,C_{x^{(0)}x^{(0)}}(t,t')\,G(t')\,,
\label{2.7}
\eeq
and for a GW interferometer $G(t)=(m/4)\, L\,\ddot{h}(t)$, where $m/4$ is the
reduced mass of the antisymmetric mode of motion of the four
arm-cavity mirrors (see Secs.~\ref{identify} and III~B). [Note that each mirror has mass $m$.] 

Henceforth, we shall assume that both the probe and the detector have
time independent
Hamiltonians, i.e.\ both $\widehat{H}_{\cal D}$ and $\widehat{H}_{\cal P}$
are time independent.
In this case, as shown in the Appendix, the susceptibilities
that appear in Eqs.~(\ref{2.4})--(\ref{2.6}) depend only on
$t-t'$. By transforming them into the Fourier domain, denoting by
$h(\Omega)$ the Fourier transform of $h(t)$ and introducing
the Fourier-domain susceptibility
\beq
\label{RAB}
R_{AB}(\Omega)\equiv\frac{i}{\hbar}\int_{0}^{+\infty}d\tau\,e^{i\Omega\tau}\,
C_{AB}(0,-\tau)\,,
\eeq
we derive 
\bea
\label{2.8}
\widehat{Z}^{(1)}(\Omega)&=&\widehat{Z}^{(0)}(\Omega)+R_{ZF}(\Omega)\,\widehat{x}^{(1)}(\Omega)\,, \\
\label{2.9}
\widehat{F}^{(1)}(\Omega)&=&\widehat{F}^{(0)}(\Omega)+R_{FF}(\Omega)\,\widehat{x}^{(1)}(\Omega)\,, \\
\label{2.10}
\widehat{x}^{(1)}(\Omega)&=& \widehat{x}^{(0)}(\Omega)+L\,h(\Omega) + R_{xx}(\Omega)\,
\widehat{F}^{(1)}(\Omega)\,.
\eea
Here and below, to simplify the notation we denote 
$R_{ZF} \equiv R_{Z^{(0)}F^{(0)}}$, $R_{FF} \equiv R_{F^{(0)}F^{(0)}}$, 
$R_{xx} \equiv R_{x^{(0)}x^{(0)}}$.
By solving Eqs.~(\ref{2.8})--(\ref{2.10}) for
the full-evolution operators in terms of the free-evolution ones, we finally get:
\bea
\label{2.11}
\widehat{x}^{(1)}(\Omega)&=&\frac{1}{1-R_{xx}(\Omega)\,R_{FF}(\Omega)}\,
\left[\widehat{x}^{(0)}(\Omega)+L\,h(\Omega) +R_{xx}(\Omega)\,\widehat{F}^{(0)}(\Omega)\right]\,,  \\
\label{2.12}
\widehat{F}^{(1)}(\Omega)&=&\frac{1}{1-R_{xx}(\Omega)\,R_{FF}(\Omega)}\,
\left[\widehat{F}^{(0)}(\Omega)+
R_{FF}(\Omega)\,\left(\widehat{x}^{(0)}(\Omega)+L\,h(\Omega)\right)\right]\,,  \\
\label{2.13}
\widehat{Z}^{(1)}(\Omega)&=& \widehat{Z}^{(0)}(\Omega)+
\frac{R_{ZF}(\Omega)}{1-R_{xx}(\Omega)\,R_{FF}(\Omega)}\,
\left[\widehat{x}^{(0)}(\Omega)+ L\,h(\Omega) + R_{xx}(\Omega)\,\widehat{F}^{(0)}(\Omega)\right]\,.
\eea
Let us point out that if the kernel relating the full-evolution 
operators to the free-evolution ones, i.e. \
$1/(1-R_{xx}\,R_{FF})$, contains poles both in the lower 
\emph{and in the upper} complex plane [with our definition 
of Fourier transform given by Eq.~(\ref{a23})],  then by applying the
standard inverse Fourier transform to Eqs.~(\ref{2.11})--(\ref{2.13}), 
we get that  $\widehat{x}^{(1)}(t)$, $\widehat{F}^{(1)}(t)$ and $\widehat{Z}^{(1)}(t)$  
depend on the gravitational-wave field and the free-evolution  
operators $\widehat{x}^{(0)}(t)$, $\widehat{F}^{(0)}(t)$ and $\widehat{Z}^{(0)}(t)$ 
both in the past \emph{and in the future}. However, 
these are not the correct solutions for the real motion. 
This situation is a very common one in physics and engineering (it occurs for example 
in the theory of linear electronic networks \cite{control} and the 
theory of plasma waves \cite{plasma}), and the cure for it is well known:
in order to obtain the (correct) full-evolution operators $\widehat{x}^{(1)}(t)$, 
$\widehat{F}^{(1)}(t)$ and $\widehat{Z}^{(1)}(t)$ 
that only depend on the past, {\it we have to alter 
the integration contour in the inverse-Fourier transform, going above 
all the poles in the complex plane}. [In the language of plasma physics 
we have to use the {\it Landau contours}.] 
This procedure, which can be justified rigorously using 
Laplace transforms \cite{laplace}, makes 
$\widehat{x}^{(1)}(t)$, $\widehat{F}^{(1)}(t)$ and $\widehat{Z}^{(1)}(t)$ 
for many systems (including LIGO-II interferometers) 
infinitely sensitive to driving forces in the 
infinitely distant past. The reason is simple and well known 
in other contexts: such quantum-measurement systems possess instabilities, 
which can be deduced from the homogeneous solutions 
of Eqs.~(\ref{2.11})--(\ref{2.13}), whose 
eigenfrequencies are given by the equation $1-R_{xx}(\Omega)\,R_{FF}(\Omega) =0$.
The zeros of the equation $1-R_{xx}(\Omega)\,R_{FF}(\Omega) =0$ are generically complex 
and for unstable systems they have positive imaginary parts, corresponding to 
homogeneous solutions that grow exponentially toward the future.

\subsection{Conditions defining a linear measurement system in terms of
susceptibilities}
\label{subsec2.3}

As we pointed out in Sec.~\ref{subsec2.1},
in order to be identified as the output of the measurement system,
the observable $\widehat{Z}$ should satisfy $[\widehat{Z}(t_1),\widehat{Z}(t_2)]=0,
\;\forall\,t_1,\,t_2$, i.e.\ the condition of {\it simultaneous measurability}.
In that section, we have
also proved the equivalence between this condition and the
condition that any external coupling to the measurement system
through $\widehat{Z}$ should not change the evolution 
of $\widehat{Z}$ itself. In the following we shall take advantage
of this equivalence: By imagining that we couple the
linear measurement system to some external system through
$\widehat{Z}$ and by looking at (possible) changes
in $\widehat{Z}$'s evolution, we shall obtain a set of conditions for the susceptibilities involving
$\widehat{Z}$.

Let us first restrict ourselves to the simplest possible external 
coupling, $\widehat{V}=-\widehat{Z}\,{\cal E}$,  
where ${\cal E}$ is a classical external force.  
The total Hamiltonian (\ref{2.1}) becomes
\beq
\label{extraction}
\widehat{H}
= \left[\left(\widehat{H}_{\cal P}-\widehat{x}\,G\right) + \widehat{H}_{\cal D} \right]
- \widehat{x}\,\widehat{F}-\widehat{Z}\,{\cal E}
=\left[\left(\widehat{H}_{\cal P}-\widehat{x}\,G\right)
+ \left(\widehat{H}_{\cal D}-\widehat{Z}\,{\cal E}\right) \right]
- \widehat{x}\,\widehat{F}\,.
\eeq
To derive the equations of motion for the Hamiltonian
(\ref{extraction}) we apply the procedure
used in Sec.~\ref{subsec2.2} to deduce the equations of motion 
for the Hamiltonian (\ref{2.1}).
First, we consider $(\widehat{H}_{\cal P}-\widehat{x}\,G)$
and $(\widehat{H}_{\cal D}-\widehat{Z}\,{\cal E})$
as zeroth order Hamiltonians and
relate the operators $\widehat{Z}^{(1)}_{\rm pert}$,
$\widehat{F}^{(1)}_{\rm pert}$ and $\widehat{x}^{(1)}_{\rm pert}$,
which evolve under the full Hamiltonian (\ref{extraction}),
to the operator $\widehat{x}^{(G)}$,
which evolves under the Hamiltonian  
$(\widehat{H}_{\cal P}-\widehat{x}\,G)$,
and the operators $\widehat{Z}^{({\cal E})}$ and $\widehat{F}^{({\cal E})}$,
evolving under the Hamiltonian $(\widehat{H}_{\cal D}-\widehat{Z}\,{\cal E})$,
\bea 
\label{2.2qm21}
\widehat{Z}^{(1)}_{\rm pert}(t)&=&\widehat{Z}^{({\cal E})}_{\rm pert}(t)+
\frac{i}{\hbar}\int_{-\infty}^{t}dt'\,C_{Z^{({\cal E})}
F^{({\cal E})}}(t,t')\,\widehat{x}^{(1)}_{\rm pert}(t')\,,\\
\label{2.3qm21}
\widehat{F}^{(1)}_{\rm pert}(t)&=&\widehat{F}^{({\cal E})}_{\rm pert}(t)+
\frac{i}{\hbar}\int_{-\infty}^{t}dt'\,
C_{F^{({\cal E})}F^{({\cal E})}}(t,t')\,\widehat{x}^{(1)}_{\rm pert}(t')\,, \\
\label{2.4qm21}
\widehat{x}^{(1)}_{\rm pert}(t)&=&\widehat{x}^{(G)}(t)+
\frac{i}{\hbar}\int_{-\infty}^{t}dt'\,C_{x^{(G)}x^{(G)}}(t,t')\,\widehat{F}^{(1)}
_{\rm pert}(t')\,.
\eea
Second, we relate the operators $\widehat{x}^{(G)}$,
$\widehat{Z}^{({\cal E})}$ and $\widehat{F}^{({\cal E})}$ to
the operators  $\widehat{x}^{(0)}$, $\widehat{Z}^{(0)}$
and $\widehat{F}^{(0)}$ which evolve under $\widehat{H}_{\cal P}$
and $\widehat{H}_{\cal D}$: 
\bea
\label{2.2qm22}
\widehat{Z}^{({\cal E})}_{\rm pert}(t)&=&\widehat{Z}^{(0)}(t)+ \frac{i}{\hbar}
\int_{-\infty}^{t}dt'\,C_{Z^{(0)} Z^{(0)}}(t,t')\,{\cal E}(t')\,, \\
\label{2.3qm22}
\widehat{F}^{({\cal E})}_{\rm pert}(t)&=&\widehat{F}^{(0)}(t)+
\frac{i}{\hbar}\int_{-\infty}^{t}dt'\,C_{F^{(0)} Z^{(0)}}(t,t')\,{\cal
E}(t')\,,\\
\label{2.4qm22}
\widehat{x}^{(G)}(t)&=&\widehat{x}^{(0)}(t) +
\frac{i}{\hbar}\int_{-\infty}^{t}dt'\,C_{x^{(0)} x^{(0)}}(t,t')\,G(t')\,.
\eea
Noticing that
$\widehat{Z}^{({\cal E})}_{\rm pert}$, $\widehat{F}^{({\cal E})}_{\rm pert}$
and $\widehat{x}^{(G)}$
differ from $\widehat{Z}^{(0)}$, $\widehat{F}^{(0)}$ and $\widehat{x}^{(0)}$
only by time dependent C-numbers, we obtain the following relations:
$C_{Z^{({\cal E})}F^{({\cal E})}}(t,t')=C_{Z^{(0)}F^{(0)}}(t,t')$,
$C_{F^{({\cal E})}F^{({\cal E})}}(t,t')=C_{F^{(0)}F^{(0)}}(t,t')$ and
$C_{x^{(G)}x^{(G)}}(t,t')=C_{x^{(0)}x^{(0)}}(t,t')$. Then, by
inserting Eqs.~(\ref{2.2qm22})--(\ref{2.4qm22}) into
Eqs.~(\ref{2.2qm21})--(\ref{2.4qm21}), we deduce the equations of motion 
of $\widehat{Z}$, $\widehat{F}$ and $\widehat{x}$
under the Hamiltonian (\ref{extraction}):
\bea
\widehat{Z}^{(1)}_{\rm pert}(t)&=&\widehat{Z}^{(0)}(t)+
\frac{i}{\hbar}\int_{-\infty}^{t}dt'\,
\left[C_{Z^{(0)} Z^{(0)}}(t,t')\,{\cal E}(t')+
C_{Z^{(0)}F^{(0)}}(t,t')\,\widehat{x}^{(1)}_{\rm pert}(t')\right]\,,
\label{2.2qm2final} \\
\widehat{F}^{(1)}_{\rm pert}(t)&=&\widehat{F}^{(0)}(t)+
\frac{i}{\hbar}\int_{-\infty}^{t}dt'\,
\left[C_{F^{(0)} Z^{(0)}}(t,t')\,{\cal E}(t')+
C_{F^{(0)}F^{(0)}}(t,t')\,\widehat{x}^{(1)}_{\rm pert}(t')\right]\,,
\label{2.3qm2final} \\
\widehat{x}^{(1)}_{\rm pert}(t)&=&\widehat{x}^{(0)}(t)+
\frac{i}{\hbar}\int_{-\infty}^{t}dt'\,C_{x^{(0)}x^{(0)}}(t,t')\,
\left[G(t')+\widehat{F}^{(1)}_{\rm pert}(t')\right]\,.
\label{2.4qm2final}
\eea
{}From Eqs.~(\ref{2.2qm2final})--(\ref{2.4qm2final}) we infer
that there are two ways the external force ${\cal E}$ can influence
the evolution of $\widehat{Z}^{(1)}_{\rm pert}$:
(i) ${\cal E}$ can affect $\widehat{Z}^{(1)}_{\rm pert}$ directly, through
the first term in the bracket of Eq.~(\ref{2.2qm2final}),
unless $C_{Z^{(0)} Z^{(0)}}(t,t')=0$ for all $t>t'$
(and thus for all pairs of $t$ and $t'$); and
(ii) ${\cal E}$ can influence the evolution of
$\widehat{Z}^{(1)}_{\rm pert}$ indirectly, affecting the
evolution of $\widehat{F}^{(1)}_{\rm pert}$ [first term
in the bracket of Eq.~(\ref{2.3qm2final})], and through it
the evolution of $\widehat{x}^{(1)}_{\rm pert}$ and
$\widehat{Z}^{(1)}_{\rm pert}$ [second terms
in the brackets of Eqs.~(\ref{2.4qm2final}), (\ref{2.2qm2final})],
unless $C_{F^{(0)} Z^{(0)}}(t,t')=0$ for all $t>t'$.

Now we are ready to deduce the conditions that 
must be satisfied in order that the evolution of
$\widehat{Z}$ not be changed by the external coupling ${\cal E}$.
In principle the two ways ${\cal E}$ affects the evolution of
$\widehat{Z}$ may cancel each other. However, noticing the fact that case (i)
does not depend on the probe (only $C_{Z^{(0)} Z^{(0)}}$ matters), but case (ii) does 
($C_{x^{(0)} x^{(0)}}$ also matters),
we see that the cancellation will not always occur
if we assume that, {\em whatever probe the detector is coupled to}, 
$\widehat{Z}$ always corresponds to the output of the
measurement process.
Thus {\it both} conditions must be satisfied: $C_{Z^{(0)} Z^{(0)}}=0$ 
and $C_{F^{(0)} Z^{(0)}}=0$. 

This argument for both conditions can be made more clear by assigning
an ``effective mass'' $\mu$ to the probe and
consider a continuous family of probes labeled by $\mu$ (for
interferometers the family of probes are the family of
mirrors with different masses). The susceptibility
of the coordinate $\widehat{x}$ depends on the effective mass as
\beq
\label{effmass}
C_{x^{(0)}x^{(0)}} \propto \frac{1}{\mu}\,,
\eeq
which simply says that the probe's response to external forces
decreases as its effective mass increases. Because $\widehat{Z}^{(0)}$
and $\widehat{F}^{(0)}$ are operators evolving
under the free Hamiltonian of the detector, they do not
depend on $\mu$. Now consider two cases: First, 
the limiting case of $\mu \rightarrow \infty$. Then
$C_{x^{(0)}x^{(0)}}\rightarrow0$ and from Eq.~(\ref{2.4qm2final})
we get $\widehat{x}^{(1)}_{\rm pert}(t)
= \widehat{x}^{(0)}(t)$. As a consequence, ${\cal E}$ affects
the evolution of $\widehat{Z}^{(1)}_{\rm pert}$ only through the first
term in the bracket of Eq.~(\ref{2.2qm2final}) [see case (i) above],
unless $C_{Z^{(0)} Z^{(0)}}(t,t')=0$ for all pairs of $t$ and $t'$.
Second, consider the case of finite mass $\mu$, and then conclude that ${\cal E}$ will affect
the evolution of $\widehat{Z}^{(1)}_{\rm pert}$ only through the second
term in the bracket of Eq.~(\ref{2.2qm2final}) [see case (ii) above],
unless $C_{F^{(0)} Z^{(0)}}(t,t')=0$ for all $t>t'$.

In conclusion we have found that if, whatever the probe is, $\widehat{Z}$ always corresponds 
to the output of the linear measurement device, then 
the following conditions must be satisfied
\bea
\label{2.14}
{\rm LQM}:
\left\{
\begin{array}{rllll}
C_{Z^{(0)} Z^{(0)}}(t,t')&
\equiv [\widehat{Z}^{(0)}(t),\widehat{Z}^{(0)}(t') ]&
=0&
\quad\quad&
\forall\,t,t'\, \\
C_{F^{(0)} Z^{(0)}}(t,t')&
\equiv [\widehat{F}^{(0)}(t),\widehat{Z}^{(0)}(t') ] &
=0&
\quad\quad&
\forall\,t>t'\,.
\end{array}
\right.
\eea
In the frequency domain these conditions read
\beq
\label{2.14FD}
R_{ZZ}(\Omega)=0=R_{FZ}(\Omega)\,.
\eeq

It is possible to show that LQM [Eqs.~(\ref{2.14})] are also 
sufficient conditions for the simultaneous measurability condition 
(\ref{commute}) be satisfied independently of the probe's nature;
imagine coupling our linear measurement system
to an external system
with an arbitrary Hamiltonian $H_{\rm EXT}$ via
a generic coupling $V(\widehat{Z}, \widehat{\cal E})$,
$\widehat{\cal E}$ being an external observable, and check whether 
the evolution of $\widehat{Z}$ is affected by this coupling. The check 
can be achieved by writing the total Hamiltonian as
\beq
\widehat{H}=
\left[
\left(\widehat{H}_{\cal P}-\widehat{x}\,G\right)
+\left(\widehat{H}_{\cal D}-\widehat{Z}\,\widehat{\cal E}+\widehat{H}_{\rm EXT}\right)
\right]
-\widehat{x}\,\widehat{F}\,,
\eeq
and re-doing all the steps followed earlier in this section. It is
helpful to notice that the evolutions of $\widehat{Z}$ and $\widehat{F}$
under
$\widehat{H}_{\cal D}-\widehat{Z}\widehat{\cal E}+\widehat{H}_{\rm EXT}$
are the same as those under $\widehat{H}_{\cal D}$, once the condition
LQM, or Eqs.~(\ref{2.14}), is satisfied. The result, after a long calculation 
is that conditions (\ref{2.14}) are sufficient to guarantee that the evolution 
of $\widehat{Z}$ is unaffected by the coupling. 
The technical details of the proof are left as an exercise for the reader.

\subsection{Effective description of measurement systems}
\label{subsec3.2}

It is common to normalize the output observable $\widehat{Z}$
to unit signal --- e.g., in the case of GW interferometer, it is common 
to set to unity the coefficient in front
of the (classical) observable $L\,h$ we want to measure 
so the normalized output $\widehat{Z}$ has the form:
\begin{equation}
\label{normalized}
\widehat{\cal O}=\widehat{\cal N}+L\,h\,,
\end{equation}
where $\widehat{\cal N}$ is the so-called {\it signal-referred} quantum noise.
The observable $\widehat{\cal O}$ can be easily deduced in the
frequency domain by renormalizing Eq.~(\ref{2.13}),
\bea
\widehat{\cal O}(\Omega) &=&
\frac{1-R_{xx}(\Omega)\,R_{FF}(\Omega)}{R_{ZF}(\Omega)}\,\widehat{Z}^{(1)}(\Omega) \nonumber \\
&=& \frac{\widehat{Z}^{(0)}(\Omega)}{R_{ZF}(\Omega)}+
R_{xx}(\Omega)\left[\widehat{F}^{(0)}(\Omega)-R_{FF}(\Omega)\,
\frac{\widehat{Z}^{(0)}(\Omega)}{R_{ZF}(\Omega)}\right ]+
\widehat{x}^{(0)}(\Omega)+L\,h(\Omega) \,,
\eea
that is
\beq
\label{2.15}
\widehat{\cal O}(\Omega) = \widehat{\cal Z}(\Omega)+R_{xx}(\Omega)\,\widehat{\cal
F}(\Omega) +\widehat{x}^{(0)}(\Omega)+L\,h(\Omega)\,.
\eeq
Here we have introduced two linear observables
$\widehat{\cal Z}$ and $\widehat{\cal F}$ 
defined in the Hilbert space ${\cal H}_{\cal D}$ of the detector,
\beq
\widehat{\cal Z}(\Omega)\equiv\frac{\widehat{Z}^{(0)}(\Omega)}{R_{ZF}(\Omega)}\,,
\quad \quad \widehat{\cal F}(\Omega)\equiv
\widehat{F}^{(0)}(\Omega)-R_{FF}(\Omega)\,
\frac{\widehat{Z}^{(0)}(\Omega)}{R_{ZF}(\Omega)}\,.
\label{2.16}
\eeq
In the time domain the output observable $\widehat{\cal O}(t)$ reads
\beq
\widehat{\cal O}(t) = \int_{-\infty}^{+\infty}dt'\,K(t-t')\,\widehat{Z}^{(1)}(t')
\label{Ot1}
\eeq
where
\beq
{K}(t)=
\int_{-\infty}^{+\infty}
\frac{1-R_{xx}(\Omega)\,R_{FF}(\Omega)}{R_{ZF}(\Omega)}
\,e^{-i\,\Omega\,t}\,\frac{d\Omega}{2 \pi}\,.
\eeq
Thus
\beq
\widehat{\cal O}(t)
 = \widehat{\cal Z}(t)+ \frac{i}{\hbar}\int_{-\infty}^t
dt'C_{x^{(0)}x^{(0)}}(t,t') \,\widehat{\cal F}(t')+\widehat{x}^{(0)}(t)+L\,h(t)\,.
\label{Ot}
\eeq
Using the two properties given by Eqs.~(\ref{a27}) of the Appendix,
and applying the conditions LQM [Eqs.~(\ref{2.14})],
we obtain the following commutation relations for the
observables $\widehat{\cal Z}(t)$ and $\widehat{\cal F}(t)$ in the Fourier domain
\beq
\label{2.18}
\left[\widehat{\cal Z}(\Omega),\widehat{\cal Z}^\dagger(\Omega')\right]
= 0= \left[\widehat{\cal F}(\Omega),
\widehat{\cal F}^\dagger(\Omega')\right]\,, \quad \quad
\left[\widehat{\cal Z}(\Omega), \widehat{\cal F}^\dagger(\Omega')\right]=-2 \pi\,i
\hbar\,\delta(\Omega-\Omega')\,,
\eeq
or in the time domain:
\footnote{~Note that, if we use the commutator of
$\widehat{\cal{Z}}$ and $\widehat{\cal{F}}$ to evaluate the
susceptibilities, we find naively that $R_{\cal{{F}{Z}}}$ and
$R_{\cal{{Z}{F}}}$ are proportional to
$\int_0^{\infty}d\tau\delta({\tau})$, which is not a well defined
quantity. However, introducing an upper cut-off $\Lambda$ in the
frequency domain we can write
$\delta{(\tau)}$ as $\delta(\tau)={\sin\Lambda\tau}/{\pi\tau}$ for 
$\Lambda\rightarrow+\infty$, which is symmetric around
the origin. With this prescription $\int_0^{+\infty}d\tau
\delta(\tau)={1}/{2}$, and the susceptibilities: 
$R_{\cal{{Z}{Z}}}=R_{\cal{{F}{F}}}=0,
R_{\cal{{F}{Z}}}=1/2, R_{\cal{{Z}{F}}}=-{1}/{2}$.}
\bea
\label{2.19a}
&& \left[\widehat{\cal Z}(t),\widehat{\cal Z}(t')\right]=
0= \left[\widehat{\cal F}(t),\widehat{\cal F}(t')\right]
\;\;\;\;\;\forall\,t,t'\,, \\
\label{2.19b}
&& \left[\widehat{\cal Z}(t),\widehat{\cal F}(t')\right]=-i \hbar\,
\delta(t-t') \;\;\;\;\;\forall\,t,t'\,.
\eea
It is interesting to notice that, because the
observables $\widehat{\cal Z}(t)$ and $\widehat{\cal F}(t)$
satisfy the commutation relations (\ref{2.19a}),
they can be regarded at different times as describing different
degrees of freedom. Moreover, because of Eq.~(\ref{2.19b}), the observables
$\widehat{\cal Z}(t)$ and $\widehat{\cal F}(t)$ can be seen
at each instant of time as the {\it canonical} momentum
and coordinate of different {\it effective} monitors (probe-detector measuring devices). 
Therefore, $\widehat{\cal Z}(t)$ and $\widehat{\cal F}(t)$ define
an infinite set of effective monitors, indexed by $t$, similar to
the successive independent monitors of von Neumann's model \cite{vonneumann} for 
quantum-measurement processes investigated by Caves, Yuen and Ozawa \cite{debate}.
However, by contrast with von Neumann's model,
the monitors defined by $\widehat{\cal Z}(t)$ and $\widehat{\cal F}(t)$
at different $t$'s
are {\it not} necessarily independent. 
They may, in fact,  have nontrivial statistical correlations, 
embodied in the relations
\beq
\langle \widehat{\cal Z}(t)\,\widehat{\cal Z}(t') \rangle  \neq {\rm const} \times \delta(t-t')\,,
\;\;
\langle \widehat{\cal F}(t)\,\widehat{\cal F}(t') \rangle  \neq {\rm const}\times \delta(t-t')\,,
\;\;
\langle \widehat{\cal Z}(t)\,\widehat{\cal F}(t') \rangle  \neq {\rm const}\times \delta(t-t')\,,
\eeq
where ``$\langle \quad \rangle$'' denotes the expectation value in 
the quantum state of the system. These correlations can be built up
automatically by the internal dynamics of the detector -- for example
they are present in LIGO-type GW interferometers \cite{KLMTV00,BC1,BC2}.

Let us now comment on the origin of the various terms appearing
in Eq.~(\ref{Ot}):

\begin{itemize}
\item The first term $\widehat{\cal Z}(t)$
describes the quantum fluctuations in the monitors' readout variable
[see also Eq.~(\ref{2.16})] which are independent of the probe.
In particular, $\widehat{\cal Z}$ does not depend on the effective mass
$\mu$ of the probe. Henceforth, we refer to $\widehat{\cal Z}$
as the {\it effective} output fluctuation.
For an interferometer, the quantum noise embodied in $\widehat{\cal Z}$ is
the well-known shot noise.
\item The second term in Eq.~(\ref{Ot}) is the effective response
of the output at time $t$ to the monitor's back-action force 
at earlier times $t' < t$. Since $C_{x^{(0)}x^{(0)}} \propto {1}/{\mu}$
this part of the output depends on the effective mass of the probe.
For GW interferometers the back action is caused by
radiation-pressure fluctuations acting on the four
arm-cavity mirrors. In the following we refer to
$\widehat{\cal F}$ as the {\it effective} back-action or
radiation-pressure force. The noise embodied in $\widehat{\cal F}$
is called the back-action noise.  [In the case of GW
interferometers, it is also called the radiation-pressure noise, since
the back-action is just the radiation-pressure force.]
\item The third term in Eq.~(\ref{Ot}) is the
free-evolution operator of the probe's coordinate.
In principle, this is also a noise term. However, in many cases
the free-evolution of the probe coordinate is confined to a
certain uninteresting frequency range, so if we make measurements outside this range,
the noise due to the free evolution of the probe will not affect the
measurement.  We shall see in
Sec.~\ref{sec3.1} that this will be the case for GW interferometers,
as has been pointed out and discussed at length by Braginsky, Gorodetsky, Khalili,
Matsko, Thorne and Vyatchanin (BGKMTV for short)~\cite{BGKMTV00}.
\item The last term in Eq.~(\ref{Ot}) is the displacement
induced on the probe by the classical observable we want to measure.
\end{itemize}

Within the effective description of the measurement's renormalized output 
[Eq.~(\ref{Ot})], it is instructive to analyze
how the simultaneous measurability condition 
$[\widehat{\cal O}(t_1),\widehat{\cal O}(t_2)]=0 \quad
\forall t_1,t_2$, is enforced by the probe-detector interaction. 
To evaluate explicitly the commutation relations of
the observable $\widehat{\cal O}$, we notice that
in Eq.~(\ref{Ot}) the first two terms always commute
with the third term, because they belong to
the two different Hilbert spaces ${\cal H}_{\cal D}$ and ${\cal
H}_{\cal P}$. The other terms give
\bea
\left[\widehat{\cal O}(t_1),\widehat{\cal O}(t_2)\right]
=&&
\left[ \widehat{\cal Z}(t_1)+ \frac{i}{\hbar}\int_{-\infty}^{t_1}
  dt'_1 C_{x^{(0)}x^{(0)}}(t_1,t'_1) \,\widehat{\cal
  F}(t'_1),\quad\widehat{\cal Z}(t_2)+ \frac{i}{\hbar}\int_{-\infty}^{t_2}
  dt'_2 C_{x^{(0)}x^{(0)}}(t_2,t'_2) \,\widehat{\cal
  F}(t'_2)\right]  \nonumber \\
&& + \left[\widehat{x}^{(0)}(t_1),\widehat{x}^{(0)}(t_2)\right]\,.
\label{commutatorseparate}
\eea
Hence, the two-time commutator of $\widehat{\cal O}(t)$ is
the sum of two terms: the first term depends solely on detector observables,
while the second term is just the two-time commutator of the free-probe
coordinate $\widehat{x}^{(0)}(t)$.  Using the commutation relations of 
$\widehat{\cal Z}(t)$ and $\widehat{\cal F}(t)$ given by 
Eqs.~(\ref{2.19a}), (\ref{2.19b}) it is straightforward to deduce 
that in Eq.~(\ref{commutatorseparate}) the detector commutator 
exactly cancels the probe commutator. 
This clean cancellation is a very interesting property of probe-detector kinds of 
quantum-measurement systems and has been recently pointed out and discussed 
at length by BGKMTV in Ref.~\cite{BGKMTV00}.

\section{Dynamics of signal recycled interferometers: Equations of Motion}
\label{sec3}
In this section we investigate the dynamics of a SR interferometer,
showing that it is a  probe-detector linear quantum-measurement
device as defined and investigated in Sec.~\ref{sec2}.

\begin{figure}[ht]
\begin{center}
\begin{tabular}{cc}
\hspace{-0.4cm}
\epsfig{file=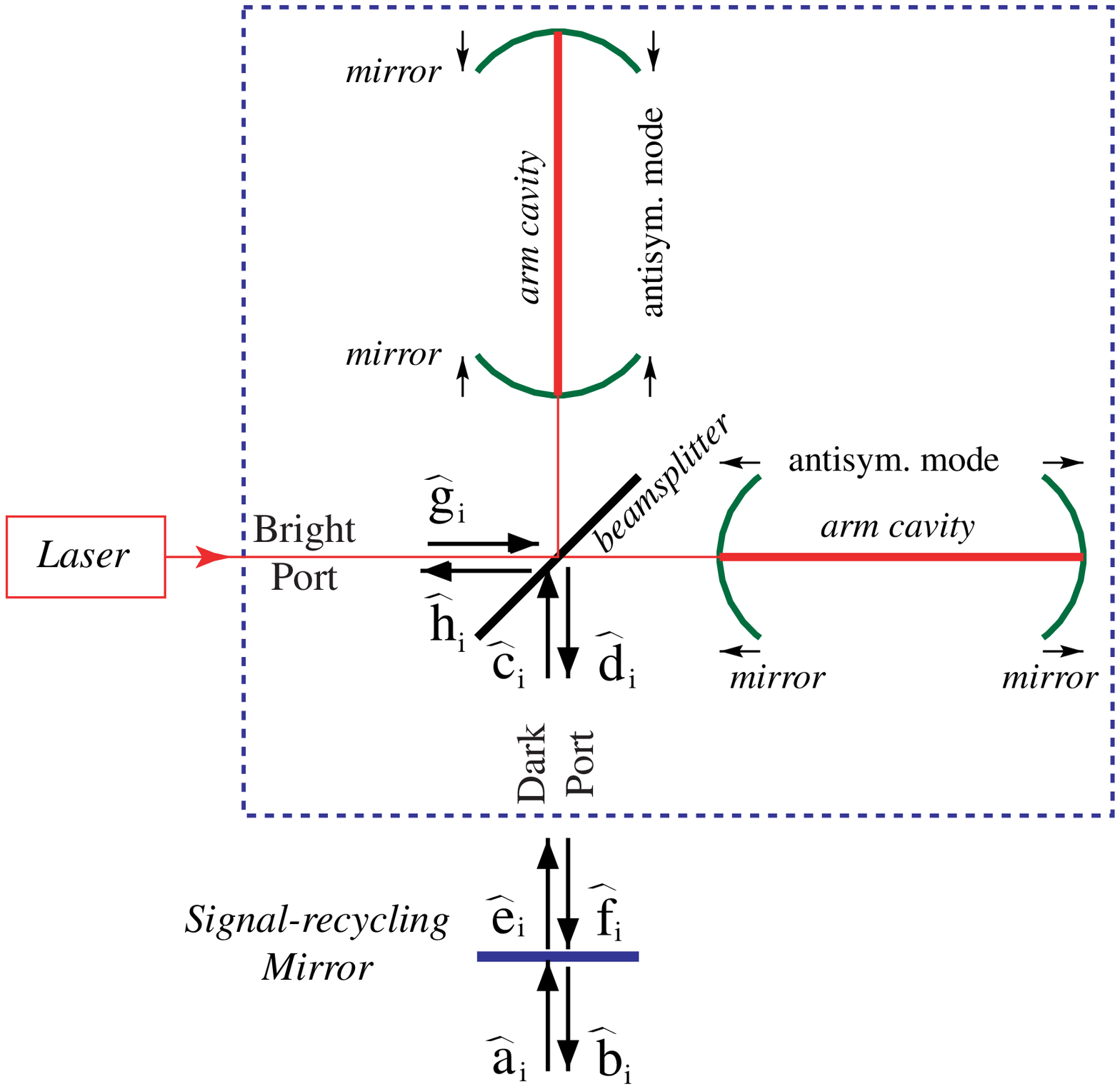,width=0.5\textwidth}&
\epsfig{file=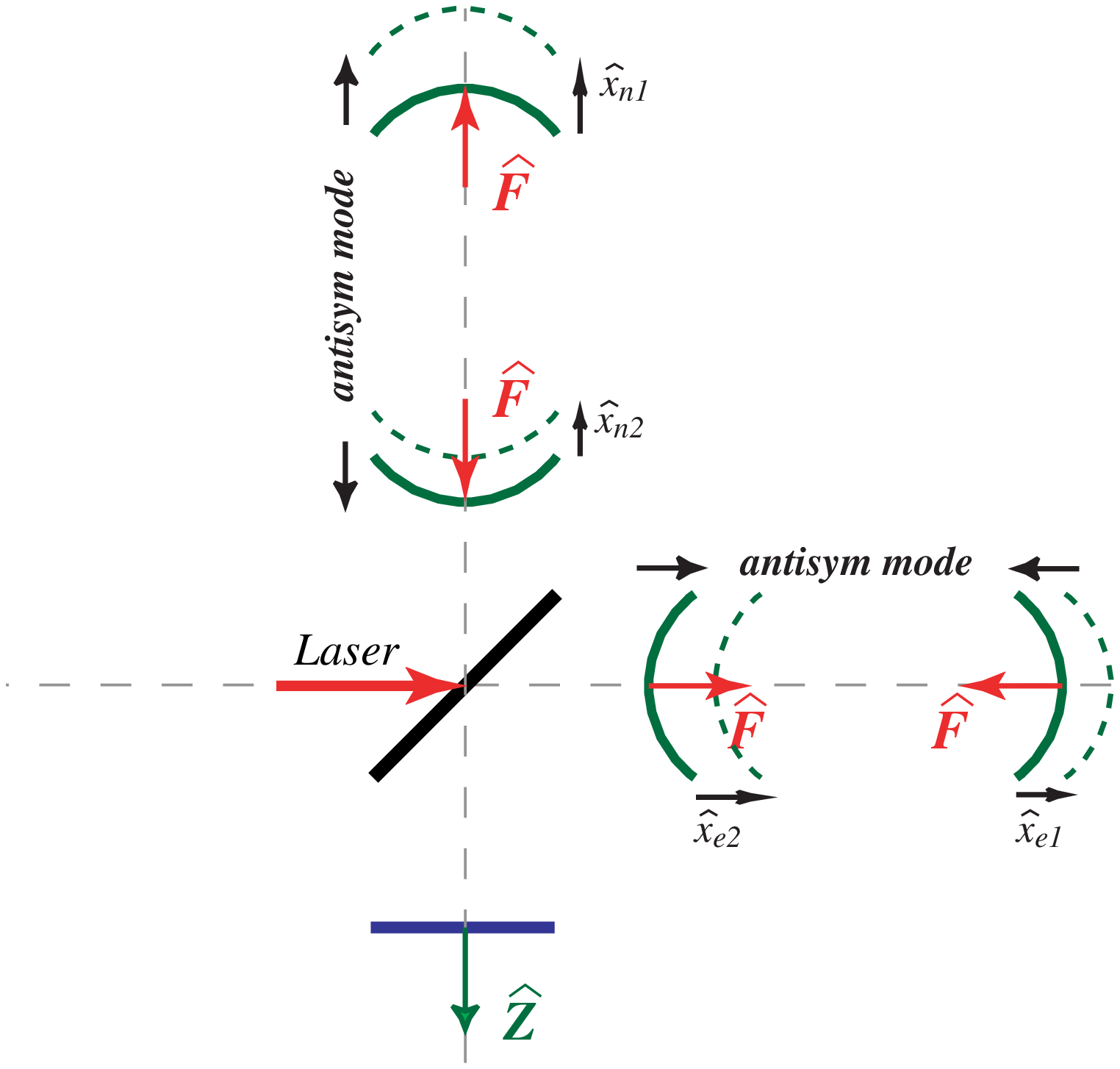,width=0.5\textwidth}
\end{tabular}
\vskip 0.2truecm
\caption{\sl On the left panel we draw a SR interferometer, showing  
the antisymmetric mode of mirror motion (marked by arrows), the  
dark-port and SR optical fields $\widehat{a}_i, \ldots ,\widehat{f}_i$ 
and the bright-port fields $\widehat{g}_i,\widehat{h}_i$, $i=1,2$. 
The conventional-interferometer optical scheme is contained inside the dashed box. 
In the right panel we identify the variables, 
$\widehat{x} \equiv 
\widehat{x}_{\rm antisym} = (\widehat{x}_{n1}-\widehat{x}_{n2})-(\widehat{x}_{e1}-\widehat{x}_{e2})$, 
$\widehat{Z}$ and $\widehat{F}$, describing the 
dynamics of the SR interferometer. } 
\label{FigSRIFO}
\end{center}
\end{figure}

\subsection{Identifying the dynamical variables and their interactions} 
\label{identify}

In gravitational-wave interferometers composed of equal-length arms (the optical 
configuration adopted by LIGO/VIRGO/GEO/TAMA), laser interferometry is used to 
monitor the displacement of the antisymmetric mode 
of the four arm-cavity mirrors induced by the passage 
of a gravitational wave (see Fig.~\ref{FigSRIFO}).

Recently Kimble, Levin, Matsko, Thorne and Vyatchanin (KLMTV for short)
\cite{KLMTV00} described a conventional (LIGO-I type) interferometer 
using a full quantum mechanical approach 
(see the optical scheme inside the dashed box in the left 
panel of Fig.~\ref{FigSRIFO}).
KLMTV \cite{KLMTV00} showed (as has long been known \cite{PCW93}) 
that in this kind of interferometer 
the antisymmetric mode of motion of the four arm-cavity mirrors 
and the dark-port sideband fields ($\widehat{c}_i$ and $\widehat{d}_i$ 
\footnote{~Here $\widehat{a}_i$, $\widehat{b}_i$,
$\widehat{c}_i$, \ldots with $i=1,2$ stand for the two quadrature operators of the
electromagnetic field. This formalism was developed by
Caves and Schumaker \cite{CS85}, adopted by KLMTV \cite{KLMTV00} and the 
authors \cite{BC1,BC2}.} in Fig.~\ref{FigSRIFO}) are 
decoupled from other degrees of freedom, i.e.\
from other modes of motion of the four arm-cavity mirrors and from 
the bright-port sideband fields ($\widehat{g}_i$ and $\widehat{h}_i$ 
in Fig.~\ref{FigSRIFO}). As a consequence, the
dynamics relevant to the output signal and the corresponding noise are 
described only by the antisymmetric mode of motion of the four arm-cavity mirrors 
and the dark-port sideband fields (see Appendix B of KLMTV \cite{KLMTV00} for details).
This result remains valid for SR interferometers \cite{BC1,BC2}:
we only need to include in the analysis all the optical fields inside
the SR cavity, such as $\widehat{c}_i$, $\widehat{d}_i$,
$\widehat{e}_i$ and $\widehat{f}_i$ [but not $\widehat{g}_i$ or $\widehat{h}_i$], 
and those outside the SR cavity, 
such as $\widehat{a}_i$ and $\widehat{b}_i$.

The coordinate of the antisymmetric mode of motion is defined by KLMTV 
[see Fig.~3 and Eq.~(12) of Ref.~\cite{KLMTV00},  
and the right panel of Fig.~\ref{FigSRIFO} in our paper] as:
\beq
\widehat{x}_{\rm antisym}
\equiv (\widehat{x}_{n1}-\widehat{x}_{n2})-(\widehat{x}_{e1}-\widehat{x}_{e2})\,,
\eeq
and we identify it with the dynamical variable $\widehat{x}$ 
introduced in Sec.~\ref{subsec2.2} [see Eq.~(\ref{x1})].
The output of the detector can be constructed from two
independent output observables, the two quadratures 
$\widehat{b}_1$ and $\widehat{b}_2$ of the outgoing electromagnetic 
field immediately outside the SR mirror (see the left panel of Fig.~\ref{FigSRIFO}). 
If a homodyne-detection read-out scheme is implemented, then the output is a linear 
combination of the two quadratures, that is
\beq
\label{quadrature}
\widehat{b}_{\zeta}\equiv
\sin\zeta\,\widehat{b}_1+
\cos\zeta\,\widehat{b}_2\,,\quad\zeta={\rm const}\,,
\eeq
which is a generic quadrature field. 
\footnote{~Rigorously speaking, the output is the photocurrent, which
in the homodyne detection scheme is almost precisely proportional 
to the output quadrature field, but not quite so; see Ref.~\cite{BGKMTV00} and
the Appendix of Ref.~\cite{BC2} for more discussion on this point.}
We thus identify the dynamical variable $\widehat{Z}$ introduced 
in Sec.~\ref{subsec2.2} [Eq.~(\ref{z1})] as:
\beq
\label{identifyZ}
\widehat{Z}_{\zeta}\equiv\widehat{b}_{\zeta}\,.
\eeq
In particular, when $\zeta=\pi/2$ and $\zeta=0$ we have
$\widehat{Z}_1\equiv\widehat{b}_1$ and $\widehat{Z}_2\equiv\widehat{b}_2$.

The radiation-pressure force acting on the arm-cavity mirrors, and 
coupled to the antisymmetric mode, can be directly related to the dark-port
quadrature fields. This result was explicitly derived in Appendix B 
of KLMTV \cite{KLMTV00}. As a foundation for subsequent calculations, 
we shall summarize the main steps 
of their derivation: 
The force acting on each arm-cavity mirror is $2 W/c$, where $W$ is the power circulating 
in each arm cavity,
which is proportional to the square of the amplitude
of the electric field propagating toward the mirror.
In the arm cavities, the electric field can be decomposed 
into two parts: the carrier and the sideband fields. Introducing 
the  carrier amplitude $C$ and the sideband quadrature operators
$\widehat{s}_{1,2}$, we have 
\beq
\label{Eincavity}
\widehat{E}(t)=C \cos \omega_0 t +
\cos \omega_0 t \,
\left[\int_0^{+\infty} \frac{d\Omega}{2\pi}\, e^{-i\,\Omega
t} \,\widehat{s}_1 + {\rm h.c.} \right]
+ \sin \omega_0 t\,
\left[\int_0^{+\infty} \frac{d\Omega}{2\pi}\, e^{-i\,\Omega
t}\, \widehat{s}_2 + {\rm h.c.} \right]\,,
\eeq
where h.c.\ stands for Hermitian conjugate.
(Note that by writing the carrier field as $C \cos\omega_0 t$,
we have adopted the convention used by KLMTV \cite{KLMTV00}.) 
Taking the square of $\widehat{E}(t)$, we obtain
\bea
\widehat{E}^2(t) &=& {[\rm DC\;component]} +
{[\rm high\; frequency\; component\; (> \! \omega_0)]} \nonumber  \\
&+& {C}\,\left[\int_0^{+\infty} \frac{d\Omega}{2\pi}\, e^{-i\,\Omega
t}\, \widehat{s}_1 + {\rm h.c.} \right]
+ (\rm quadratic\;terms\;in\;\widehat{s}_1,\,\widehat{s}_2)\,,
\label{E2}
\eea
where we have used the fact that in the integral $\Omega < \omega_0$. 
The DC and $\omega_0 \sim 10^{15} \,{\rm sec}^{-1}$ components are not in 
the detection band of GW interferometers, 
$10\,{\rm Hz}\leq \Omega/2\pi \leq 10^4\,{\rm Hz}$; in practice 
they will be  counteracted by control
systems. We also ignore the quadratic terms in Eq.~(\ref{E2}), since they are much
smaller than the linear terms. Thus, modulo a factor of
proportionality, we obtain in the Fourier domain the 
following expression for the radiation-pressure force acting on each mirror: 
\beq
\label{FRP}
\widehat{F}_{\rm RP}(\Omega) \propto C \,\widehat{s}_1(\Omega)\,. 
\eeq
As shown in Appendix B of Ref.~\cite{KLMTV00}, the in-cavity quadrature 
field $\widehat{s}_1$ is a combination of the incoming quadratures 
from both the dark and the bright ports. However,
the contribution from the bright-port fields do not couple to
the antisymmetric mode, so the force acting on the antisymmetric mode is
due only to the incoming fields from the dark port.
More specifically, in Sec. 4 of Appendix B of Ref.~\cite{KLMTV00}, KLMTV related the in-cavity carrier 
amplitude $C$ and the sideband quadrature $\widehat{s}_1$ (which they denoted by $\widehat{j}_1$ 
\footnote{~We ignore the effect of the arm-cavity optical losses, thus in this case 
the quadratures $\widehat{j}_i$ and $\widehat{k}_i$ in 
Ref.~\cite{KLMTV00} are equal.}) to the input carrier amplitude and ingoing
dark-port quadrature $\widehat{c}_1$ (which they denoted by $\widehat{a}_1$). 
Although they did not give the explicit expression we need here for $\widehat{F}_{\rm RP}$, 
it is straightforward to recover it. Using the arrows indicated
in the right panel  of Fig.~\ref{FigSRIFO} as positive directions,
we find\footnote{~This result can be obtained from Eq.~(B21) of KLMTV
\cite{KLMTV00} using the fact that $\widehat{x}_{\rm BA} = -{4}/{m \Omega^2}\,\widehat{F}_{\rm RP}$.
Since in this paper we ignore optical losses, in Eq.~(B21) we can replace  
$\beta_*$ and ${\cal K}_*$ by $\beta$ and ${\cal K}$ 
and ignore the noise operator $\widehat{n}_1$.}
\beq
\widehat{F}_{\rm RP}=\sqrt{\frac{2 I_0 \hbar \omega_0}
{\left(\Omega^2+\gamma^2\right)L^2}}\,e^{i\,\beta}\,\widehat{c}_1,
\eeq
where $\omega_0$ is the carrier laser frequency, $I_0$ is the carrier light power 
entering the
beamsplitter, $2\beta=2\arctan{{\Omega}/{\gamma}}$ is the net phase
gained by the sideband frequency $\Omega$ while in the arm cavity,
$\gamma = Tc/4L$ is the half bandwidth of the arm cavity
($T$ is the power transmissivity of the input mirrors and
$L$ is the length of the arm cavity). We identify the force 
$\widehat{F}_{\rm RP}$ with the dynamical variable $\widehat{F}$ 
introduced in Sec.~\ref{subsec2.2} [see Eq.~(\ref{f1})]:
\beq
\label{identifyF}
\widehat{F}\equiv\widehat{F}_{\rm RP}
=\sqrt{\frac{2 I_0 \hbar
\omega_0}{\left(\Omega^2+\gamma^2\right)L^2}}\,e^{i\,\beta}\,\widehat{c}_1\,.
\eeq
Applying Newton's law to the four mirrors, we deduce
\beq
m\,\ddot{\widehat{x}}=4 \widehat{F} +{\rm other\;forces}\,,
\eeq
where ``other forces'' refer to forces not due to the optical-mechanical
interaction, e.g., the force due to the gravitational wave and thermal forces.
By identifying the reduced mass of the antisymmetric mode as $m/4$, 
we obtain that the coupling term in the total Hamiltonian 
(\ref{2.1}) is $-\widehat{x}\,\widehat{F}$.
[The reduced mass coincides with the effective mass of the probe $\mu$ 
introduced in Sec.~II.]

Note that, by assuming the four forces acting on the arm-cavity mirrors are 
equal,  
we have made the approximation used by KLMTV \cite{KLMTV00}
of disregarding the motion of the mirrors during the 
light's round-trip time (quasi-static approximation).
\footnote{~The description of a SR interferometer beyond the quasi-static approximation 
\protect\cite{MMPDHV,R00} introduces nontrivial corrections to the back-action force,  
proportional to the power transmissivity of the input arm-cavity mirrors. 
Since the power transmissivity expected for LIGO-II is very small, we expect a small 
modification of our results, but an explicit calculation is 
much needed to quantify this effect.}

\subsection{Free evolutions of test mass and optical field}
\label{sec3.1}

In this section we derive the dynamics of the free probe and the detector, i.e.\ 
that of the antisymmetric mode of motion of the arm-cavity mirrors
when there is no light in the arm cavities, and that of the optical fields when the
arm-cavity mirrors are held fixed. The full, coupled dynamics will be
discussed in the following section.

The mirror-endowed test masses are suspended from seismic isolation
stacks and have free oscillation frequency $\sim \!1\,{\rm Hz}$. However, since
we are interested in frequencies above $\sim \! 10\, {\rm Hz}$ 
(below these frequencies the seismic noise is dominant),
we can approximate the antisymmetric-mode
 coordinate as the coordinate of a free particle with (reduced) mass
$m/4$ --- as is also done by KLMTV \cite{KLMTV00}. Hence, its free evolution is given by
\beq
\widehat{x}^{(0)}(t)=\widehat{x}_s+\frac{4}{m}\,\widehat{p}_s\, t\,,
\label{freeprobe}
\eeq
where $\widehat{x}_s$ and $\widehat{p}_s$ are the Schr\"odinger
operators of the canonical coordinate and momentum of the mode.
Inserting Eq.~(\ref{freeprobe}) into Eqs.~(\ref{susc}), 
(\ref{RAB}) and using the usual commutation relations 
$[\widehat{x}_s, \widehat{p}_s] = i \hbar$, it is straightforward to derive 
\beq
\label{Rxx}
R_{xx} = - \frac{4}{m\Omega^2}\,.
\eeq
As discussed in detail by BGKMTV \cite{BGKMTV00}, 
since at frequencies below $\laq \, 10$ Hz the data will be filtered out,
the free evolution observable $\widehat{x}^{(0)}(t)$, whose Fourier component 
has support only at zero frequency (in a real interferometer it has support 
at the pendulum frequency $\sim \! 1\,{\rm Hz}$), does not contribute to the 
output noise. For this reason, henceforth, we shall disregard 
the free-evolution observable $\widehat{x}^{(0)}(t)$ in the equations of 
motion describing the dynamics of GW interferometers. 

Concerning the free detector (the light with fixed mirrors), we can solve its 
dynamics by expressing the various quantities in terms
of the quadrature operators of the input field at the SR mirror,  
$\widehat{a}_i$, $i=1,2$ (see Fig.~\ref{FigSRIFO}).
For LIGO-II the input field will be in the vacuum state.
All the quantum fluctuations affecting the output optical 
field $\widehat{b}_i$ are due to the vacuum fluctuations $\widehat{a}_i$ 
entering the interferometer from the SR mirror.

Through Eqs.~(\ref{identifyZ}), (\ref{identifyF}), 
we have already expressed $\widehat{Z}$ and $\widehat{F}$ in terms
of the quadrature fields $\widehat{b}_{\zeta}$ and $\widehat{c}_1$;  
thus we need now to relate the latter to $\widehat{a}_i$, $i=1,2$,
This can be done using Eqs.~(2.11),(2.15)--(2.19) of Ref.~\cite{BC2}, 
in the case of fixed mirrors. First, for the input-output 
relation at the beam splitter  (see Fig.~\ref{FigSRIFO}) we have
\beq
\label{prd1_2.11}
\widehat{d}_1 = \widehat{c}_1\, e^{2 i \beta}\,, \quad \quad \widehat{d}_2=
\widehat{c}_2\, e^{2 i \beta}\,,
\eeq
which is obtained from Eq.~(2.11) of Ref.~\cite{BC2}, or Eq.~(16) of Ref.~\cite{KLMTV00} 
in the limit $I_0 \rightarrow 0$ and $h\rightarrow 0$, i.e.\
when we neglect the effects of mirror motion under radiation pressure
and gravitational waves. Second, propagating the quadrature fields inside the SR cavity,  
we obtain [see Eqs.~(2.16), (2.17) of Ref.~\cite{BC2}]
\bea
\label{prd1_2.16}
\widehat{f}_1 = (\widehat{d}_1\,\cos{\phi}-\widehat{d}_2\,\sin{\phi})\,, \quad && \quad
\widehat{f}_2 = (\widehat{d}_1\,\sin{\phi}+\widehat{d}_2\,\cos{\phi})\,, \\
\label{prd1_2.17}
\widehat{e}_1 = (\widehat{c}_1\,\cos{\phi}+\widehat{c}_2\,\sin{\phi})\,, \quad && \quad
\widehat{e}_2 =(-\widehat{c}_1\,\sin{\phi}+\widehat{c}_2\,\cos{\phi})\,,
\eea
where $\phi\equiv [\omega_0 l/c]_{{\rm mod}\;2\pi}$ is the phase gained
by the carrier frequency $\omega_0$ traveling one-way in the SR cavity, and  
for simplicity we have neglected the tiny additional phase $\Phi \equiv {\Omega l}/{c}$
gained by the sideband frequency $\Omega/2 \pi$ in the SR cavity. 
[The length of the SR cavity is typically $l\sim \! 10\,{\rm m}$, hence $\Phi \ll 1$.] 
{} From the reflection/transmission 
relations at the SR mirror we derive [see Eqs.~(2.18), (2.19) of Ref.~\cite{BC2}]
\bea
\label{prd1_2.18}
&& \widehat{e}_1 = \tau\, \widehat{a}_1+\rho\,  \widehat{f}_1 \,, \quad \quad
\widehat{e}_2 = \tau\, \widehat{a}_2+\rho\, \widehat{f}_2 \,,\\
\label{prd1_2.19}
&& \widehat{b}_1= \tau\,\widehat{f}_1-\rho\,\widehat{a}_1\,, \quad \quad \,\,
\widehat{b}_2 = \tau\,\widehat{f}_2-\rho\,\widehat{a}_2\,,
\eea
where $\tau$ and $\rho$ are the transmissivity and reflectivity of the SR mirror, with
$\tau^2+\rho^2=1$. \footnote{~For simplicity we ignore the effects of
optical losses which were discussed in Sec.~V of Ref.~\cite{BC2}.}
Solving Eqs.~(\ref{prd1_2.11})--(\ref{prd1_2.19}) and using 
Eq.~(\ref{identifyZ}), we obtain for the free-evolution operators
\bea
\widehat{Z}^{(0)}_1(\Omega) \equiv
\left[\widehat{b}_1(\Omega)\right]_{\rm mirrors\;fixed}&=
&\frac{e^{2i\beta}}{M_0}\left \{
\left[\left(1+\rho^2 \right)\,\cos 2\phi -2 \rho\, \cos 2\beta \right]\,\widehat{a}_1
-\tau^2\, \sin 2\phi\,\widehat{a}_2 \right \}\,,
\label{3.1a}
\\
\widehat{Z}^{(0)}_2(\Omega) \equiv
\left[\widehat{b}_2(\Omega)\right]_{\rm mirrors\;fixed}&=&\frac{e^{2i\beta}}{M_0}
\left \{\tau^2\, \sin 2\phi\,\widehat{a}_1+
\left[\left(1+\rho^2\right)\,\cos 2\phi -2 \rho\, \cos 2\beta\right]\,\widehat{a}_2
\right \}\,,
\label{3.1b} \\
\left[\widehat{c}_1(\Omega)\right]_{\rm mirrors\;fixed}&=&
\frac{\tau \left [\left(1-\rho\,e^{2i\beta}\right)\,\cos\phi\,\widehat{a}_1 -
\left(1+\rho\,e^{2i\beta}\right)\,\sin\phi\,
\widehat{a}_2 \right ]}{M_0}\,
\label{c1}
\eea
where we have defined,
\bea
M_0(\Omega) &\equiv& 1 +\rho^2\,e^{4i\beta}-2\rho\,\cos 2\phi \,e^{2i\beta}\,
= \left(1+2\rho\,\cos2\phi+\rho^2\right)\,
\frac{(\Omega-\Omega_+)\,(\Omega-\Omega_-)}{(\Omega + i\gamma)^2}
\,,
\label{3.2}
\eea
and
\beq
\label{3.6}
\Omega_{\pm} = \frac{1}{1 + 2\rho\,\cos 2\phi + \rho^2}\,
\left [\pm 2\rho\,\gamma\,\sin 2\phi - i \gamma\,(1 - \rho^2)\right ] \,.
\eeq
Note that $\widehat{Z}^{(0)}_{\zeta}$ can be computed from 
Eqs.~(\ref{3.1a}), (\ref{3.1b})  by taking the linear combination
of $\widehat{Z}^{(0)}_1$ and $\widehat{Z}^{(0)}_2$, in the manner of 
Eqs.~(\ref{quadrature}), (\ref{identifyZ}). 
{}From Eqs.~(\ref{identifyF}) and (\ref{c1}) 
we obtain for the free-evolution radiation-pressure force: 
\footnote{~Note that if we take the limit $\tau \rightarrow 0$, 
$\widehat{F}^{(0)}(\Omega)$ does not go to zero but 
$\sim \delta(\Omega \pm \gamma\,\tan \phi)$. Thus the main 
contribution of the fluctuating force comes from frequencies 
close to $\Omega = \pm \gamma\,\tan \phi$, which are the optical 
resonances of the interferometer with arm-cavity mirrors fixed.}
\bea
\widehat{F}^{(0)}(\Omega)&=&
\tau\, \sqrt{\frac{2 I_0 \hbar \omega_0}
{\left(\Omega^2+\gamma^2\right)L^2}}\,\frac{e^{i\,\beta}}{M_0}
\left [\left(1-\rho\,e^{2i\beta}\right)\,\cos\phi\,\widehat{a}_1 -
\left(1+\rho\,e^{2i\beta}\right)\,\sin\phi\,
\widehat{a}_2 \right ]\,.
\label{3.3}
\eea

Using Eqs.~(\ref{3.1a}), (\ref{3.1b}) and (\ref{3.3}), and the fact
that $\zeta$ is frequency independent,
we have explicitly checked that the susceptibilities
of the free-evolution operators, $\widehat{Z}_{\zeta}^{(0)}$ and
$\widehat{F}^{(0)}$, satisfy the necessary and sufficient conditions LQM, given  
in Sec.~\ref{subsec2.3}, which define a linear quantum-measurement system 
with output $\widehat{Z}$. More specifically, using the commutation relations among 
the quadrature fields $\widehat{a}_1$ and $\widehat{a}_2$
[Eqs.~(7a), (7b) of Ref.~\cite{KLMTV00}], namely
\bea
\label{prd1_2.8}
&& [\widehat{a}_1, \widehat{a}_{2^\prime}^{\dagger}] = - 
[\widehat{a}_2, \widehat{a}_{1^\prime}^{\dagger}]=
2\pi\, i\, \delta(\Omega-\Omega^\prime)\,,
\label{prd1_com1} \\
&& [\widehat{a}_1, \widehat{a}_{1^\prime}^{\dagger}] = 0= [
\widehat{a}_1, \widehat{a}_{1^\prime}]\,,
\quad \quad [\widehat{a}_2, \widehat{a}_{2^\prime}^{\dagger}] = 0= 
[\widehat{a}_2, \widehat{a}_{2^\prime}]\,,
\label{prd1_com2}
\eea
we have derived that 
\beq
R_{Z_{\zeta}Z_{\zeta}}=0=R_{FZ_{\zeta}}\,;
\eeq
and we have also derived that
\bea
R_{FF}(\Omega)&=&\frac{ 2 I_0\,\omega_0}{L^2}\,\frac{\rho\,\sin 2\phi}{1+2\rho\, \cos 2\phi +\rho^2}\,
\frac{1}{(\Omega-\Omega_+)\,(\Omega-\Omega_-)}\,,
\label{3.4a}
\\
R_{Z_1 F}(\Omega) &=&-i\,\sqrt{\frac{2 I_0\,\omega_0}{\hbar\, L^2}}\,
\frac{\tau \sin\phi}{1+2\rho\, \cos 2\phi +\rho^2}\,
\frac{(1-\rho)\,\Omega+i(1+\rho)\,\gamma}
{(\Omega-\Omega_+)\,(\Omega-\Omega_-)}\,,
\label{3.4b} \\
R_{Z_2 F}(\Omega) &=&i\,\sqrt{\frac{2 I_0\,\omega_0}{\hbar\,L^2}}\,
\frac{\tau \cos\phi}{1+2\rho\,\cos 2\phi +\rho^2}\,
\frac{(1+\rho)\,\Omega+i(1-\rho)\,\gamma}
{(\Omega-\Omega_+)\,(\Omega-\Omega_-)} \,,
\label{3.4c}
\eea
\beq
\label{3.4d}
R_{Z_{\zeta}{F}}(\Omega) = R_{Z_1 F}(\Omega) \, \sin\zeta +  R_{Z_2
F}(\Omega) \, \cos\zeta \,.
\eeq

In actuality the commutation relations (\ref{prd1_com1}), (\ref{prd1_com2}) are
approximate expressions for $\Omega \ll \omega_0$. However, this is a good approximation since
the sideband frequency $\Omega/2\pi$ varies over the range  $10-10^4\,{\rm Hz}$, 
which is ten orders of magnitude smaller than
$\omega_0/2\pi \! \sim \! 10^{14}\,{\rm Hz}$. 
If we had used the exact commutation relations (see Caves and Schumaker \cite{CS85} or
Eqs.~(2.4), (2.5) of Ref.~\cite{BC2}), we would still have $R_{FZ_{\zeta}}=0$, 
\footnote{~It is quite straightforward to understand why $R_{FZ_{\zeta}}$ must be zero. 
In fact $\widehat{Z}_{\zeta}$ is the amplitude of an outgoing wave; thus, the 
operator $\widehat{Z}_{\zeta}$ at an earlier time 
cannot be causally correlated with $\widehat{F}$ at any later time, and 
as a consequence $[\widehat{F}^{(0)}(t_1),\widehat{Z}_{\zeta}^{(0)}(t_2)]=0$ for $t_1>t_2$.}
but we would have correction terms 
in the other susceptibilities. In particular, $R_{Z_{\zeta}Z_{\zeta}}$ would not vanish, 
but would instead be on the order of $\Omega/\omega_0$.
These issues are discussed in the Appendix of Ref.~\cite{BC2}.

Before ending this section we want to discuss the resonant
features of the free-evolution optical fields, which originally 
motivated the Signal Recycling (SR) \cite{D82,VMMB88,M88} and
Resonant Sideband Extraction (RSE) schemes \cite{MSNCSRWD93,M95,H99}.
By definition a resonance is an infinite response
to a driving force acting at a certain (complex) frequency. Mathematically, 
it corresponds to a pole of the Fourier-domain susceptibility at that
(complex) frequency. {}From Eqs.~(\ref{3.4a})--(\ref{3.4d})
we deduce that $R_{FF}$ and $R_{Z_{\zeta}F}$ have only two poles $\Omega_{\pm}$, 
given by Eq.~(\ref{3.6}), which are the two complex resonant frequencies of the free 
optical fields, Eqs.~(\ref{3.1a}), (\ref{3.1b}). The corresponding eigenmodes are of the form 
$e^{-t/\tau_{\rm decay}}\,e^{- i\, \Omega_{\rm osc}\,t}$, with oscillation frequency
\beq
\Omega_{{\rm osc}\pm}={\Re}(\Omega_{\pm})=
\pm \frac{2\rho\,\gamma\,\sin 2\phi}{1 + 2\rho\,\cos 2\phi + \rho^2}\,,
\label{Omegaosc}
\eeq
and decay time
\beq
\tau_{\rm decay}=
-\frac{1}{{\Im}(\Omega_{\pm})}=\frac{1 + 2\rho\,\cos 2\phi + \rho^2}{\gamma\,(1-\rho^2)}\,.
\label{taudecay}
\eeq

This oscillation frequency and decay time give information 
on the frequency of perturbations to which the optical fields are most sensitive, 
and on the time these perturbations last in the 
interferometer before leaking out. 
Let us focus on several limiting cases:
\begin{itemize}
\item[(i)] For $\rho=0$, i.e.\ the case of a conventional (LIGO-I type) 
of interferometer,
we have $\Omega_{\rm osc}=0$ and $\tau_{\rm decay}=1/\gamma$. Thus, there is no oscillation, 
while the decay time $1/\gamma$ of the entire interferometer is just the
storage time of the arm cavity.
\item[(ii)] For $\rho\rightarrow 1$, i.e.\ when the SR optical system is nearly closed, 
we have $\Omega_{\rm osc}=\pm \gamma \tan\phi$ and $\tau_{\rm decay}\rightarrow + \infty$,
which corresponds to a pure oscillation.
Noticing that for sideband fields with frequency $\Omega/2 \pi$, the phase gained
in the arm cavity is $2\beta=2\arctan{\Omega/\gamma}$ and the phase gained during a
round trip in the SR cavity is $2\phi=2\omega_0 l/c$,
we obtain that $\Omega_{\rm osc}$ is just the frequency
at which the total round-trip phase in the entire cavity (arm cavity  + SR cavity) is $2\pi n$,  
with $n$ an integer.
\item[(iii)] For $0<\rho <1$ and $\phi=0$, we get $\Omega_{\rm osc}=0$ and
$\tau_{\rm decay}=({1+\rho})/[{\gamma\,(1-\rho)}] > {1}/{\gamma}$. This
is the so-called tuned SR configuration \cite{D82,VMMB88,M88},
where the sideband fields remain in the inteferometer for a time longer than the storage time 
of the arm cavities [cf. (i)].
\item[(iv)] For $0<\rho <1$ and $\phi=\pi/2$, we get $\Omega_{\rm osc}=0$ and
$\tau_{\rm decay}=({1-\rho})/[{\gamma\,(1+\rho)}] < {1}/{\gamma}$. This is the
so-called tuned RSE configuration
\cite{MSNCSRWD93,M95,H99}, where the sideband fields remain in the interferometer 
for a time shorter than the storage time of the arm cavities [cf. (i)].
\end{itemize}

\subsection{Coupled evolution of test mass and optical field:
ponderomotive rigidity}
\label{sec3.2}

In Sec.~\ref{subsec2.2} we have solved the equations of motion for a generic 
quantum-measurement device by expressing the
full-evolution operators in terms of the free-evolution operators
[see Eqs.~(\ref{2.11})--(\ref{2.13})].
Using the free-evolution optical-field operators (\ref{3.1a}), (\ref{3.1b}) and (\ref{3.3})
and the optical-field susceptibilities (\ref{3.4a})--(\ref{3.4d}), along with the
susceptibility of the antisymmetric mode (\ref{Rxx}),
\footnote{~As was discussed at the beginning of Sec.~\ref{sec3.1},  
the free-evolution operator $\widehat{x}^{(0)}$ describing 
the antisymmetric mode is irrelevant since it will
be filtered out during the data analysis.}
we can now obtain the full evolution of the antisymmetric mode 
$\widehat{x}^{(1)}$ and that of the output optical 
field $\widehat{Z}^{(1)}_{\zeta}$ for a SR interferometer. 
In Ref.~\cite{BC2}, we evaluated the output quadrature
fields by a slightly different method, introduced by KLMTV
\cite{KLMTV00}. However, the approach followed in this paper
provides the output field in a more straightforward way,  
and gives a clearer understanding of the interferometer dynamics.
Moreover, we think this method 
is more convenient when the optical configuration of the interferometer 
is rather complex. 

We start by investigating the interaction between the probe and the detector.   
The equations that couple the various quantities $\widehat{x}$,
$\widehat{F}$ and $\widehat{Z}$ 
are [Eqs.~(\ref{2.8})--(\ref{2.10})]:
\bea
\widehat{Z}_{\zeta}^{(1)}(\Omega)&=&\widehat{Z}_{\zeta}^{(0)}(\Omega)+R_{Z_{\zeta}F}(\Omega)\,
\widehat{x}^{(1)}(\Omega)\,,
\label{3.9} \\
\label{3.7}
\widehat{F}^{(1)}(\Omega)&=&\widehat{F}^{(0)}(\Omega)+R_{FF}(\Omega)\,
\widehat{x}^{(1)}(\Omega)\,, \\
\widehat{x}^{(1)}(\Omega)&=&
R_{xx}(\Omega)\,[G(\Omega)+ \widehat{F}^{(1)}(\Omega)]\,. \label{3.8}
\eea
In these equations, we have made explicit the dependence on
the gravitational force $G(\Omega) = -(m/4)\,\Omega^2\,h(\Omega)$ [see
also Eq.~(\ref{2.7})] and have neglected the free
evolution operator $\widehat{x}^{(0)}$ (see the discussion at the
beginning of Sec.~\ref{sec3.1}).

Equation (\ref{3.8}) is the equation of motion of the antisymmetric mode
under the GW force $G$ and the radiation-pressure force
$\widehat{F}$, with response function $R_{xx}$.
Equations~(\ref{3.9}) and (\ref{3.7}) are the equations of motion of
the optical fields $\widehat{Z}_{\zeta}$ and $\widehat{F}$
under the modulation of the antisymmetric mode of motion of the four arm-cavity 
mirrors $\widehat{x}$, with response functions $R_{Z_{\zeta}F}(\Omega)$
and $R_{FF}(\Omega)$, respectively.

The optical-mechanical interaction in a conventional interferometer 
($\rho=0$ and $\phi=0$) was analyzed by KLMTV in Ref.~\cite{KLMTV00}. 
Here we summarize only the main features. Inside the arm cavity the 
electric field is [see Eq.~(\ref{Eincavity})]
\bea
\widehat{E}(t) 
&\propto& C \cos \omega_0 t +\widehat{S}_1(t) \cos \omega_0 t
+ \widehat{S}_2(t) \sin \omega_0 t \,, \nonumber \\
&\approx& C \left [1+\frac{\widehat{S}_1(t)}{C} \right ]
\cos \left [\omega_0 t - \frac{\widehat{S}_2(t)}{C} \right ]\,,
\label{S}
\eea
with 
\beq
\widehat{S}_j(t)= \int_0^{+\infty} \frac{d\Omega}{2\pi}\, e^{-i\,\Omega\, t}\,\widehat{s}_j + 
{\rm h.c.}\,,\quad \quad j=1,2\,,
\eeq
where in Eq.~(\ref{S}) we have assumed that the sideband amplitudes are much smaller than the
carrier amplitude. From Eq.~(\ref{S}) we infer that the sideband fields
$\widehat{S}_1$ and $\widehat{S}_2$ modulate the amplitude and the phase
of the carrier field. 
If the arm-cavity mirrors are not moving, then it is easy to deduce 
that $\widehat{b}_1 \propto \widehat{s}_1 \propto \widehat{a}_1$ and 
$\widehat{b}_2 \propto \widehat{s}_2 \propto \widehat{a}_2$
(see Fig.~\ref{FigSRIFO}). Thus, given our conventions 
for the quadratures, we can refer to $\widehat{s}_1$, $\widehat{a}_1$ and $\widehat{b}_1$ 
as amplitude quadratures, and $\widehat{s}_2$, $\widehat{a}_2$ and $\widehat{b}_2$ 
as phase quadratures in the present case of a conventional interferometer. 
When the arm-cavity mirrors move, their motion 
modulates the phase of the carrier field, pumping part of it into the phase quadrature
$\widehat{S}_2(t)$, and thus into $\widehat{b}_2$ [see Appendix B 
of Ref.~\cite{KLMTV00}, especially Eq.~(B9a)]. 
As a consequence $R_{Z_2 F}\ne0$ but $R_{Z_1 F}=0$.  
On the other hand, the radiation-pressure force
acting on the arm-cavity mirrors is determined by the amplitude modulation
$\widehat{S}_1(t)$ and does not respond to the 
motion of the arm-cavity mirrors; thus $R_{FF}=0$.

Let us now analyze a SR interferometer. As pointed out above, 
the antisymmetric mode of motion of the arm-cavity mirrors, $\widehat{x}$, 
only appears in the phase quadrature $\widehat{d}_2$. 
[Note that now $\widehat{c}_i$ and $\widehat{d}_i$ take the place 
of $\widehat{a}_i$ and $\widehat{b}_i$ in the above analysis of 
conventional interferometers.]
Schematically,
\beq
\left( \begin{array}{c} \widehat{c}_1 \\ \widehat{c}_2 \end{array}\right)
\stackrel{\rm arm \atop cavity}
{\longrightarrow}
e^{i\,({\rm phase})}
\left( \begin{array}{c} \widehat{c}_1 \\ \widehat{c}_2 \end{array}\right)
+
\left( \begin{array}{c} 0 \\ \widehat{x} \end{array}\right)
\Leftrightarrow
\left( \begin{array}{c} \widehat{d}_1 \\ \widehat{d}_2 \end{array}\right)\,.
\label{Armcavity}
\eeq
Because of the presence of the SR mirror, part of the field coming out from the beamsplitter  
is reflected by the SR mirror and fed back into the arm cavities. 
Due to the propagation inside the SR cavity, the outgoing amplitude/phase 
quadrature fields at the beamsplitter, $\widehat{d}_{1,2}$, get rotated 
[see Eqs.~(\ref{prd1_2.16}), (\ref{prd1_2.17})].
Moreover, whereas part of the light leaks out from the SR mirror,
contributing to the output field, some vacuum fields
leak into the SR cavity from outside [see Eqs.~(\ref{prd1_2.18}), (\ref{prd1_2.19})].
When the light reflected by the SR mirror, along with the vacuum fields that have leaked in, 
reaches the beamsplitter again, the rotation angle is $2\phi\,$. 
Schematically, we can write 
\beq
\left(\begin{array}{l} \widehat{d}_1 \\ \widehat{d}_2 \end{array} \right)
\stackrel{\rm SR \atop cavity}{\longrightarrow}
\rho
\left(\begin{array}{cc} \cos 2\phi & -\sin 2\phi \\ \sin 2\phi & \cos 2\phi \end{array}\right)
\left(\begin{array}{l} \widehat{d}_1 \\ \widehat{d}_2 \end{array} \right)
+
\tau
\left({\rm vacuum\;fields \atop from\;outside}\right)
\Leftrightarrow
\left(\begin{array}{l}\widehat{c}_1 \\ \widehat{c}_2
\end{array}\right)\,,
\label{SRmechanism}
\eeq
where $\rho$ and $\tau$ are the amplitude reflectivity and transmissivity of the
SR mirror.

In the particular case of $\phi=0\;{\rm or}\;{\pi}/{2}$, 
namely the {\it tuned} SR/RSE configurations \cite{M88,MSNCSRWD93,M95,H99},
the rotation matrix in Eq.~(\ref{SRmechanism}) is diagonal.
Since $\widehat{x}$ appears only in $\widehat{d}_2$ [see Eq.~(\ref{Armcavity})],
the fact that the propagation matrix is diagonal guarantees that 
$\widehat{x}$ remains only in the quadratures $\widehat{d}_2$ and $\widehat{c}_2$.
As a result, the radiation-pressure force, which is proportional
to $\widehat{c}_1$ [see Eq.~(\ref{identifyF})], is not affected by the antisymmetric 
mode of motion, and $R_{FF}=0$ [see Eq.~(\ref{3.4a})] as in
conventional interferometers. 
Moreover, since the quadratures at the beamsplitter $\widehat{d}_{1,2}$ are 
rotated by an angle of $\phi$ when they reach the SR mirror
[see Eq.~(\ref{prd1_2.16})], the information on the motion 
of the arm-cavity mirrors is contained only in the output quadrature 
$\widehat{b}_2$ for $\phi=0$ and $\widehat{b}_1$ for $\phi={\pi}/{2}$.
Therefore $R_{Z_1 F}=0$ for $\phi=0$ and $R_{Z_2 F}=0$ for
$\phi={\pi}/{2}$, as obtained directly from Eqs.~(\ref{3.4b}), (\ref{3.4c}).

For a generic configuration with $\phi\ne 0\;{\rm or}\; {\pi}/{2}$, 
which is often referred to as the \emph{detuned} case \cite{M88},
$\widehat{x}$ appears in both the quadratures $\widehat{c}_{1,2}$
as a consequence of the nontrivial rotation in Eq.~(\ref{SRmechanism}).
Thus the radiation-pressure force and
both the output quadratures respond to $\widehat{x}$, i.e.\ 
$R_{FF}\ne 0$ and $R_{Z_{\zeta}F}\ne 0$ for all $\zeta$, as can be seen from
Eqs.~(\ref{3.4a})--(\ref{3.4c}).

Before ending this section let us make some remarks. When $R_{FF}=0$, 
as occurs in conventional interferometers and the tuned SR/RSE configurations,
we infer from Eqs.~(\ref{Rxx}), (\ref{3.7}) and (\ref{3.8}) that 
\beq
-\frac{m}{4}\,\Omega^2\,\widehat{x}^{(1)}(\Omega) = 
G(\Omega) + \widehat{F}^{(0)}(\Omega)\,.
\eeq
This means that the antisymmetric mode of motion of the 
four arm-cavity mirrors behaves as a free test mass subject to the GW force 
$G(\Omega)$ and 
the fluctuating radiation-pressure force $\widehat{F}^{(0)}$. 
It is well known that for such systems the Heisenberg
uncertainty principle imposes a limiting noise spectral
density $S_h^{\rm SQL} = 8 \hbar/(m\Omega^2L^2)$ for the
dimensionless gravitational-wave signal $h(t) = \Delta L/L$ \cite{KT80}.
This limiting noise spectral density is called
the standard quantum limit (SQL) for GW interferometers, and
LIGO/VIRGO/GEO/TAMA interferometers can beat this SQL only if 
correlations among the optical fields are introduced \cite{SFD,HFD,KLMTV00,BC1,BC2}.

When $R_{FF} \ne0$, Eqs.~(\ref{Rxx}), (\ref{3.7}) and (\ref{3.8}) give
\beq
-\frac{m}{4}\,\Omega^2\,\widehat{x}^{(1)}(\Omega) = 
G(\Omega) + \widehat{F}^{(0)}(\Omega) + R_{FF}(\Omega)\,\widehat{x}^{(1)}(\Omega)\,.
\eeq
Thus the antisymmetric mode of motion of the four arm-cavity mirrors
is not only disturbed randomly by the fluctuating force
$\widehat{F}^{(0)}$, but also, and more fundamentally,  is  
subject to a linear restoring force with a
frequency-dependent rigidity (or ``spring constant'')
$K(\Omega) = -R_{FF}(\Omega) \ne 0$,
generally called a {\it ponderomotive rigidity} \cite{OB}.
This phenomenon was originally analyzed in ``optical-bar'' GW detectors 
by Braginsky, Khalili and colleagues,
where the ponderomotive rigidity affects the internal
mirror, i.e.\ an intra-cavity
meter which couples the two resonators with
end-mirror--endowed test masses \cite{OB}.
Hence, SR interferometers do not monitor the displacements
of a free test mass but instead that of a test mass subject to a force field
$\widehat{F}_{\rm res}(\Omega) = - K(\Omega)\,\widehat{x}^{(1)}(\Omega)$.
This suggests that the SQL, derived from the monitoring of a free test
mass, is irrelevant
for detuned SR interferometers. Indeed,
in Ref.~\cite{BC1,BC2} we found that there exists a region of the 
parameter space $\rho$, $\phi$ and $I_0$ for which 
the quantum noise curves can beat the SQL by roughly 
a factor of two over a bandwidth $\Delta f \! \sim \! f$.

\section{Dynamics of Signal recycled interferometers: Resonances and Instabilities}
\label{sec4}

In the previous section we have shown that in a SR interferometer 
the four arm-cavity mirrors are subject to a 
frequency dependent restoring force. Thus we expect the mirrors' motion 
be characterized by resonances and possible instabilities.
In Refs.~\cite{BC1,BC2}, we have identified those resonances by evaluating 
the input-output relation for the quadrature fields 
$\widehat{b}_i$ ($\widehat{a}_i, h$). In this section, by using   
the dynamics of the whole system composed of the optical
fields and the mirrors, we shall investigate in more detail 
the features of those resonances and instabilities.

\subsection{Physical origins of the two pairs of resonances}
\label{origin}

\begin{figure}[ht]
\begin{center}
\begin{tabular}{cc}
\epsfig{file=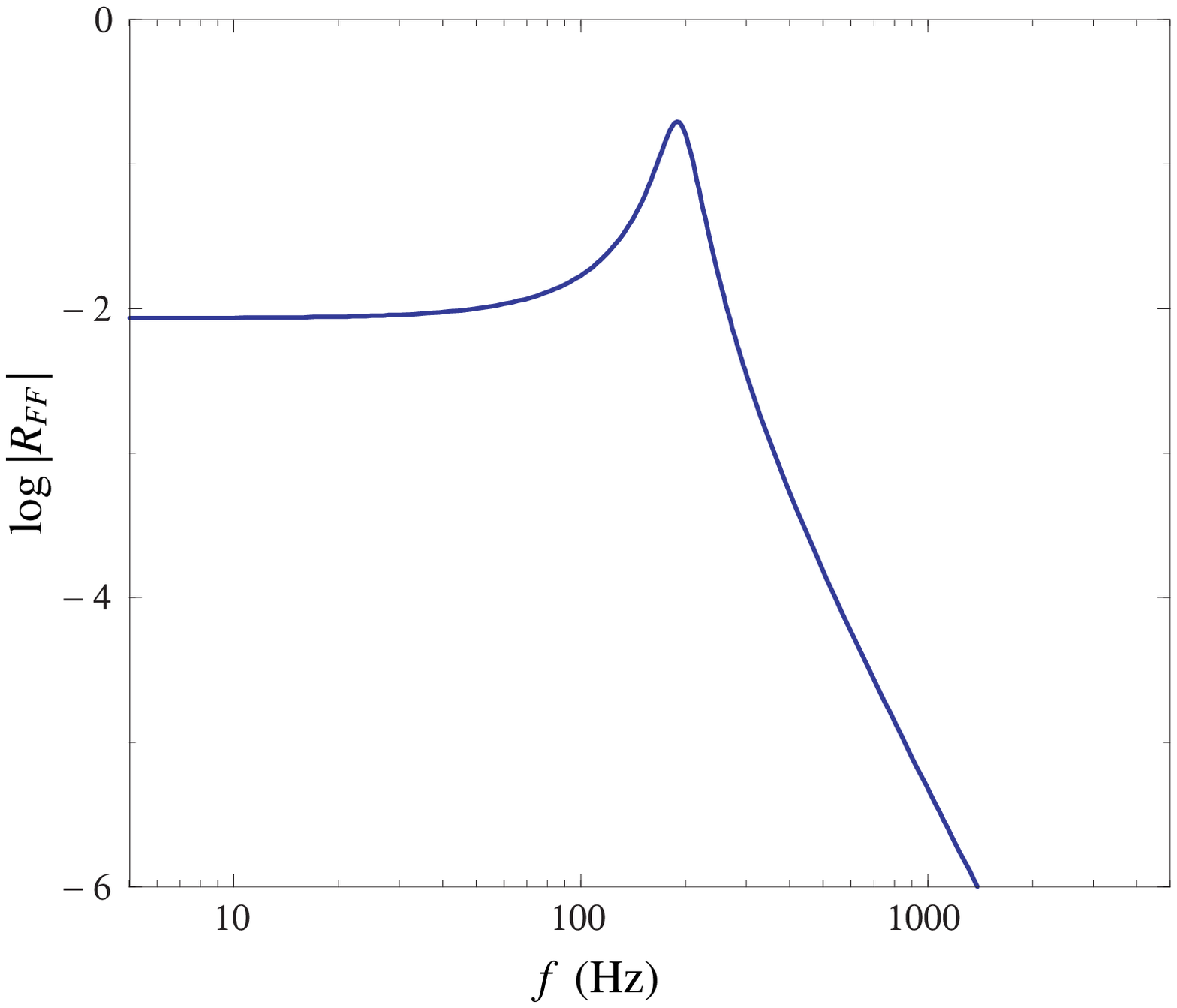,width=0.45\textwidth,height=0.4\textwidth,angle=0}
\hspace{0.02\textwidth} &
\hspace{0.02\textwidth}
\epsfig{file=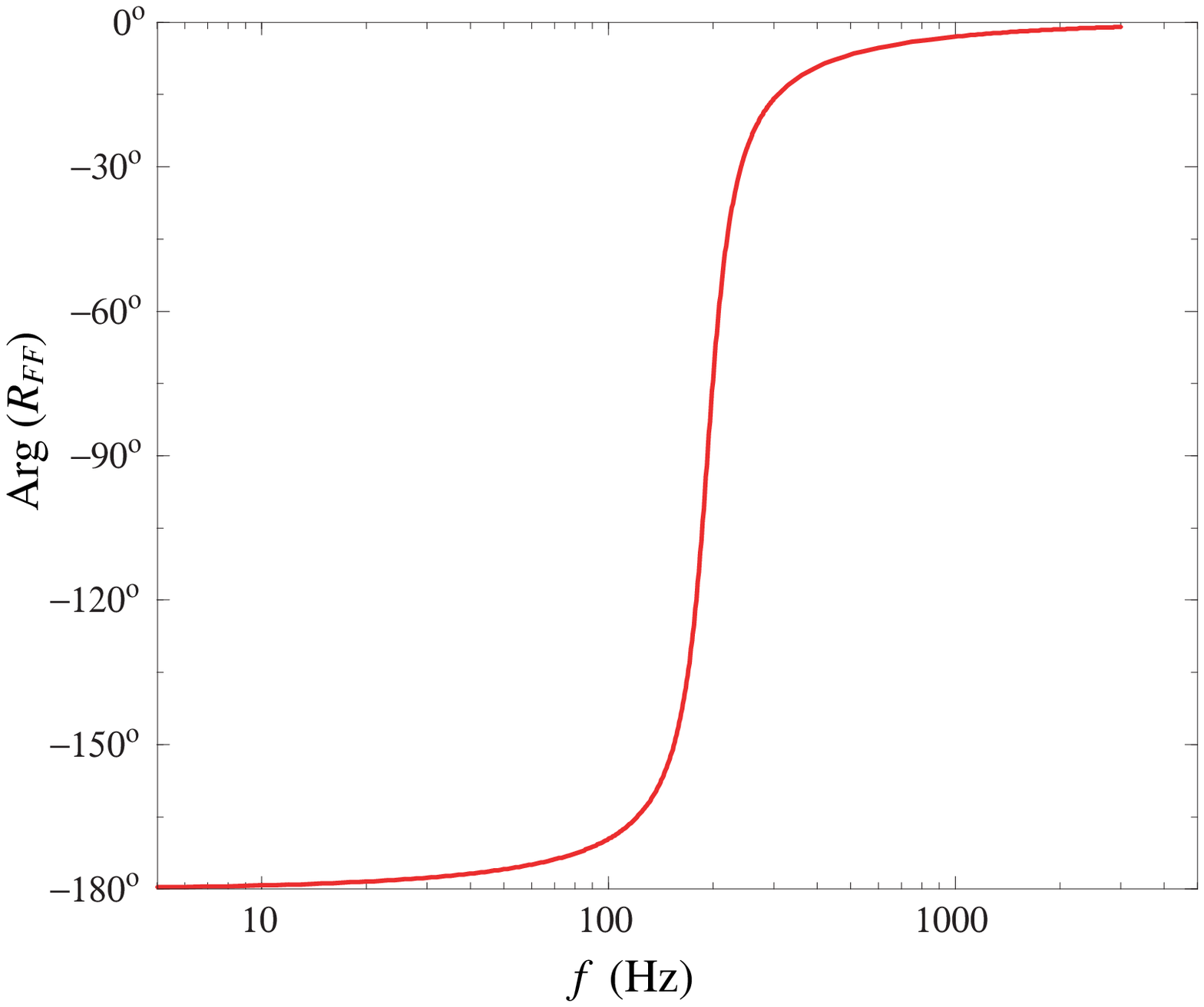,width=0.45\textwidth,height=0.4\textwidth,angle=0}
\end{tabular}
\vskip 0.2truecm
\caption{\sl Amplitude (on the left panel) and phase (on the right panel) of $R_{FF}$ 
as a function of the sideband frequency $f=\Omega/2\pi$ 
for $\phi={\pi}/{2}-0.47$, $\rho=0.9$ and $I_0 \simeq 
10^4$ W. Note that the amplitude of $R_{FF}$ is shown
in arbitrary unit.}
\label{FigRFF}
\end{center}
\end{figure}

Let us first seek a qualitative understanding of the resonances.
In Fig.~\ref{FigRFF} we draw the amplitude and the phase of the ponderomotive 
rigidity $R_{FF}$, given by Eq.~(\ref{3.4a}), for a typical choice 
of LIGO-II parameters: $\phi={\pi}/{2}-0.47$, $\rho=0.9$ and $I_0
\simeq 10^4\,{\rm W}$.
The amplitude and phase of $R_{FF}$ resemble those  
of  the response function of a damped harmonic oscillator, 
except for the fact that the phase of $R_{FF}$ is reversed.
{} From Fig.~\ref{FigRFF} we infer that when the frequency 
$f=\Omega/2\pi$ is small, $|R_{FF}|$ is almost constant, 
while the phase is nearly $-180^{\circ}$. Thus in this frequency region   
the spring constant is approximately a constant positive number 
$\sim \! K(\Omega=0) = - R_{FF}(\Omega=0) >0$. However, $K(\Omega=0)$ is positive only if $0 <\phi<{\pi}/{2}$, 
while for ${\pi}/{2}<\phi<\pi$ the spring constant at low frequencies 
is negative. As a consequence, for $\pi/2 < \phi < \pi$, 
there is a non-oscillating instability, namely a pair of
complex-conjugate purely imaginary resonant frequencies.
[Note that because the SR-interferometer dynamics is invariant  
under the transformation $\phi\rightarrow\phi+\pi$ \cite{BC2}, we 
can restrict ourselves to $0 \leq \phi \leq \pi$.] 

For larger $f=\Omega/2\pi$, $K(\Omega)=-R_{FF}(\Omega)$ has a resonant peak centered at 
$\Omega=\Omega_{\rm osc}$, with width $\sim \! 1 / \tau_{\rm decay}$ 
[see Eqs.~(\ref{Omegaosc}), (\ref{taudecay})].

Hence, the dynamics of the system composed of the optical field and 
the arm-cavity mirrors in a SR interferometer is analogous 
to the dynamics of a massive spring, with an internal mode, attached to a 
test mass. When the test mass moves at low frequency, 
i.e.\ $\Omega \ll \Omega_{\rm osc}$, the internal
configuration of the spring has time to keep up with its motion and
it remains uniform, providing a linear restoring force which  
induces a pair of resonances  at frequencies 
$\Omega_{\rm mech} = \pm \sqrt{{4 K(\Omega\ll\Omega_{\rm osc})}/{m}}
\! \sim \!\pm \sqrt{{4 K(\Omega=0)}/{m}}$.

When the test mass moves at high frequency, the internal mode of the spring 
is excited, providing another pair of resonances to the system.
Inserting the equation of motion (\ref{3.8}) of $\widehat{x}$
and the expression for $R_{FF}$, Eq.~(\ref{3.4a}), into the equation
of motion (\ref{3.7}) of $\widehat{F}$, we obtain 
\beq
-(\Omega-\Omega_+)\,(\Omega-\Omega_-)\,\widehat{F}^{(1)}(\Omega)
= {\rm driving\;terms} +
\frac{4}{m\Omega^2}\,
\frac{ 2 I_0 \omega_0}{L^2}\,\frac{\rho\sin 2\phi}
{1+2\rho \cos 2\phi +\rho^2}\,\widehat{F}^{(1)}(\Omega)\,.
\label{F}
\eeq
In the absence of the SR mirror, i.e.\ for $\rho=0$, 
the term proportional to $\widehat{F}^{(1)}$ on the RHS of 
Eq.~(\ref{F}) vanishes, and the optical field is characterized 
by the two resonant frequencies $\Omega_{\pm}$ given by Eq.~(\ref{3.6}).
By contrast, when the SR mirror is present,  
the term proportional to $\widehat{F}^{(1)}$ on the RHS of 
Eq.~(\ref{F}) shifts the resonant frequencies away from
the values $\Omega_{\pm}$. 

\begin{figure}[ht]
\begin{center}
\epsfig{file=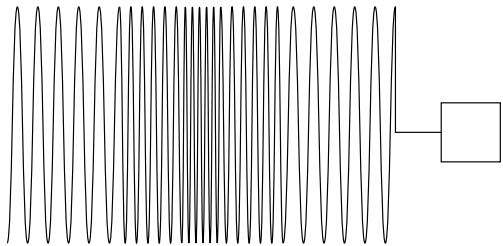,width=0.4\textwidth,angle=0}
\vspace{0.5cm}
\caption{\sl The SR-interferometer dynamics resembles the dynamics 
of a massive spring with one internal oscillation mode (and damping) attached 
to a test mass. The overall dynamical system is characterized by two pairs of 
resonances.}
\label{FigSpring}
\end{center}
\end{figure}

In conclusion, the dynamics of SR interferometers is characterized 
by two  (pairs of)  resonances with different origin: 
the (pair of)  resonances at low frequency have  a ``mechanical'' origin, coming 
from the linear restoring force due to the ponderomotive rigidity; 
the  (pair of)  resonances at higher frequency have an ``optical'' origin. 
Because of the motion of the arm-cavity mirrors 
the optical resonant frequencies get shifted
away from the free-evolution SR resonant frequencies $\Omega_{\pm}$. 
In this sense we can regard the SR interferometer as an ``optical
spring'' [See Fig.~\ref{FigSpring}].

\subsection{Quantitative investigation of the resonances}
\label{quantitative}

Equations (\ref{3.9})--(\ref{3.8}) describe the coupled evolution  
of the dynamical variables $\widehat{x}$, $\widehat{F}$ and $\widehat{Z}$:  
\bea
\label{2.11G}
\widehat{x}^{(1)}(\Omega)&=&\frac{R_{xx}(\Omega)}{1-R_{xx}(\Omega)\,R_{FF}(\Omega)}\,
\left[G(\Omega)+\widehat{F}^{(0)}(\Omega)\right]\,,  \\
\label{2.12G}
\widehat{F}^{(1)}(\Omega)&=&\frac{1}{1-R_{xx}(\Omega)\,R_{FF}(\Omega)}\,
\left[\widehat{F}^{(0)}(\Omega)+
R_{FF}(\Omega)\,R_{xx}(\Omega)G(\Omega)\right]\,,  \\
\label{2.13G}
\widehat{Z}_{\zeta}^{(1)}(\Omega)&=& \widehat{Z}^{(0)}_{\zeta}(\Omega)+
\frac{R_{Z_{\zeta}F}(\Omega)\,R_{xx}(\Omega)}{1-R_{xx}(\Omega)\,R_{FF}(\Omega)}\,
\left[G(\Omega)+
\widehat{F}^{(0)}(\Omega)\right]\,.
\eea
Let us first analyze these equations in the 
low--laser-power limit, 
which has long been considered in the literature 
for the SR/RSE schemes \cite{D82,VMMB88,M88,MSNCSRWD93,M95,H99}
and has recently been tested experimentally \cite{FHSMSLWSRWD00,M01}.
For LIGO-II \cite{GSSW99} low--laser-power limit corresponds to $I_0 \ll 10^4\,{\rm W}$. 
Using Eqs.~(\ref{3.4a})--(\ref{3.4d}),  
and the fact that $\widehat{Z}_\zeta^{(0)}$ does not depend on 
$I_0$, and $\widehat{F}^{(0)}\propto\sqrt{I_0}$  
[see Eqs.~(\ref{3.1a}), (\ref{3.1b}) and (\ref{3.3})], 
we deduce that $R_{FF}\propto I_0$ and $R_{Z_{\zeta} F}\propto \sqrt{I_0}$. 
Therefore, for very low laser power, if we restrict ourselves only to 
terms up to the order of $\sqrt{I_0}$, we 
can reduce Eq.~(\ref{2.13G}) to:
\beq
\label{2.13G_LP}
\left[
\widehat{Z}_{\zeta}^{(1)}(\Omega)\right]_{\rm low\;power}= \widehat{Z}_\zeta^{(0)}(\Omega)+
R_{Z_\zeta F}(\Omega)\,R_{xx}(\Omega)\,G(\Omega)\,,
\eeq
which says that the response of $\widehat{Z}_\zeta^{(1)}$ to the GW force $G$ 
is given by the product of $R_{xx}$, the response of $\widehat{x}$ to $G$, 
times $R_{Z_\zeta F}$, the response of $\widehat{Z}_\zeta$ to $\widehat{F}$.
Hence, for low laser power the dynamics is characterized 
by four {\it decoupled} resonant frequencies: two of them,
$\Omega^2=0$ (degenerate),
are those of the free test mass as embodied in $R_{xx}$;
the other two, $\Omega=\Omega_{\pm}$ [see Eq.~(\ref{3.6})],
are those of the free-evolution optical fields as embodied in  $R_{Z_\zeta F}$. 
As was discussed in Sec.~\ref{subsec2.2}, when the imaginary part of the
resonant frequency is negative (positive) the mode is stable
(unstable). Therefore the decoupled ``mechanical'' resonances $\Omega^2=0$ are 
marginally stable, while the decoupled ``optical'' resonances $\Omega_{\pm}$ are 
stable. [We remind the reader that $\Im(\Omega_{\pm}) <0$.]

If we increase the laser power sufficiently, the effect of the radiation pressure 
is no longer negligible, and from Eqs.~(\ref{2.11G})--(\ref{2.13G})
we derive the following condition for the resonances:
\beq
\label{rescondition}
\frac{R_{xx}(\Omega)\,R_{Z_{\zeta} F}(\Omega)}{1-R_{xx}(\Omega)\,R_{FF}(\Omega)}\rightarrow + \infty\,
\eeq
which simplifies to,
\beq
\label{3.11Mod}
\Omega^2\,(\Omega-\Omega_+)\,(\Omega-\Omega_-)
+\frac{I_0\,\gamma^3}{2 I_{\rm SQL}}\,(\Omega_+-\Omega_-)=0\,.
\eeq
In these equations we have adopted as a reference light power 
$I_{\rm SQL} \equiv {m\,L^2\,\gamma^4}/{4 \omega_0}$, introduced by 
KLMTV \cite{KLMTV00}; this is the light power at the beamsplitter 
needed by a conventional interferometer to reach
the SQL at $\Omega=\gamma$. Because of the presence 
of the term proportional to $I_0$ 
in Eq.~(\ref{3.11Mod}), $\Omega^2=0$ and
$\Omega=\Omega_{\pm}$ are no longer the resonant frequencies 
of the coupled SR dynamics.

If the laser power is not very high, we expect  the roots 
of Eq.~(\ref{3.11Mod}) to differ only slightly from the decoupled ones.
Let us then apply a perturbative analysis. 
Concerning the double roots $\Omega^2=\Omega_0^2=0$,
working at leading order in the frequency shift $\Delta \Omega_0 = \Omega - \Omega_0 = 
\Omega$, 
we derive
\beq
(\Delta \Omega_0)^2 = - \frac{I_0\,\gamma^3}{2I_{\rm SQL}} \,
\frac{(\Omega_+ - \Omega_-)}{\Omega_+\,\Omega_-}
= \frac{I_0}{I_{\rm SQL}} \,
\frac{(2\rho\,\gamma^2\,\sin 2\phi)\,(1 + 2\rho\,\cos 2 \phi +
\rho^2)}{4\rho^2\,\sin^2 2\phi + (1-\rho^2)^2}\,.
\eeq

If the SR detuning phase lies in the range $0 < \phi < \pi/2$,
then $(\Delta \Omega_0)^2$ is always positive.
Hence, at leading order, the initial double zero 
resonant frequency $\Omega^2=0$
splits into two real resonant frequencies having opposite signs and
proportional to $(I_0/I_{\rm SQL})^{1/2}\,\gamma$.
The imaginary parts of these resonant
frequencies appear only at the next to leading order, and
it turns out (as discussed later on in this section)
that they always increase  (becoming more positive) as 
$I_0/I_{\rm SQL}$ grows, generating instabilities.

If the SR detuning phase lies in the range 
$\pi/2 < \phi < \pi$, then at leading order  $(\Delta \Omega_0)^2$  
is negative, and we get two complex-conjugate purely imaginary roots. 
The system is therefore characterized by a non-oscillating instability.

Regarding the roots $\Omega=\Omega_{\pm}$, we can
expand Eq.~(\ref{3.11Mod}) with respect to $\Delta \Omega_{\pm}=
\Omega-\Omega_{\pm}$. A simple calculation gives
\beq
\Delta \Omega_{\pm} = \mp \frac{I_0\,\gamma^3}{2I_{\rm SQL}}\,\frac{1}{(\Omega_{\pm})^2}\,.
\eeq
Using Eq.~(\ref{3.6}) we find that
\bea
&& \Re (\Delta \Omega_{\pm}) = \mp \frac{I_0\,\gamma}{2I_{\rm SQL}}\,
\frac{[4\rho^2\,\sin^2 2\phi - (1-\rho^2)^2]\,(1 + 2\rho\,\cos 2 \phi +
\rho^2)^2}{[4\rho^2\,\sin^2 2\phi + (1-\rho^2)^2]^2}\,,\\
&& \Im (\Delta \Omega_{\pm})=  -\frac{I_0}{I_{\rm SQL}}\,
\frac{[2\rho\,\gamma\,\sin 2\phi\,(1-\rho^2)]\,(1 + 2\rho\,\cos 2 \phi +
\rho^2)^2}{[4\rho^2\,\sin^2 2\phi + (1-\rho^2)^2]^2}\,.
\eea
This says that, if the SR detuning phase lies in the range 
$0 < \phi < \pi/2$, then $\Im(\Delta \Omega_{\pm})$ always decreases (becoming more negative) 
as $I_0/I_{\rm SQL}$ increases.
Hence, the imaginary parts of the resonant frequencies are pushed
away from the real $\Omega$ axis, i.e.\ the system remains stable.
On the other hand, $\Re(\Delta \Omega_{\pm})$
may either increase or decrease as $I_0/I_{\rm SQL}$ grows.
If $\pi/2 < \phi < \pi$ then the imaginary parts 
become less negative as the laser power increases, so the system becomes 
less stable.
\begin{figure}[ht]
\vskip -0.4truecm
\begin{center}
\epsfig{file=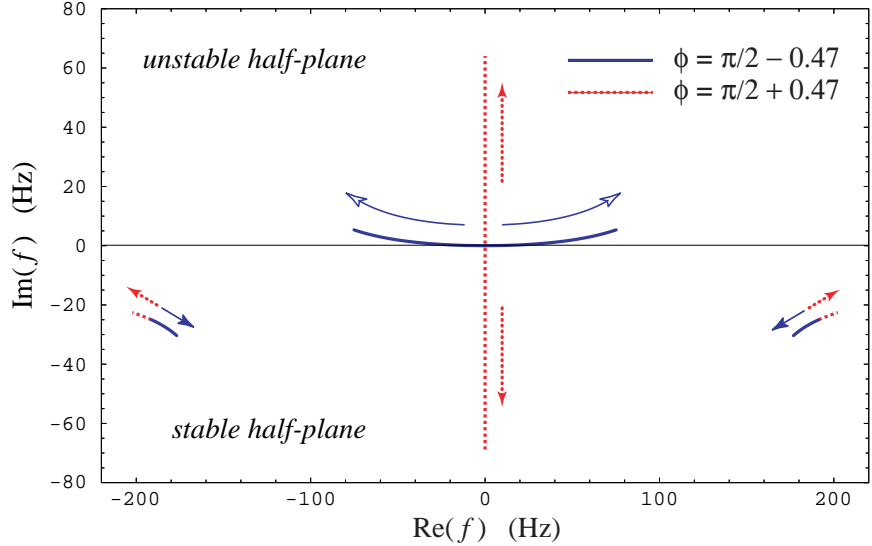,width=0.7\textwidth,angle=0}
\vskip 0.2truecm
\caption{\sl Shift of the resonances 
in a SR interferometer induced by the radiation pressure 
force as $I_0$ increases from $\sim \! 0$ up to $I_{\rm SQL}$. This figure 
is drawn for a SR mirror reflectivity $\rho = 0.9$.}
\label{FigResonance}
\end{center}
\end{figure}

Note that, although turning up the laser power drives
the optical resonant frequencies away from their nonzero values
$\Omega_{\pm}$, their changes are very small
or comparable to their original values. By contrast,
the mechanical resonant frequencies move away from zero; hence
their motion is very significant. 
In this sense, as the laser power increases, the mechanical (test-mass) 
resonant frequencies move faster than the
optical ones.
This fact can also be understood by observing that 
$\Delta \Omega_0$ is proportional to the
square root of $I_0$, while $\Delta \Omega_{\pm}$ is proportional
to $I_0$ itself. For the optical configurations of interest
for LIGO-II, we found \cite{BC2} that when we increase
the laser power from $I_0=0$ to $I_0=I_{\rm SQL}$, the optical 
resonant frequencies stay more or less close to their original values
while the mechanical ones, which start from zero at
$I_0=0$, move into the observation band of LIGO-II
as $I_0 \rightarrow I_{\rm SQL}$. 

To get a more intuitive idea of the shift in the resonant frequencies 
for high laser power, we have explored the resonant features numerically.
In Fig.~\ref{FigResonance} we plot the trajectories of the
resonant frequencies when $I_0$ varies from $\sim \!0$
to $I_{\rm SQL}$ (the arrows indicate the directions of increasing power), 
for two choices of SR parameters: $\rho=0.9$, and $\phi=\pi/2 \mp 0.47$,
for which the decoupled resonant frequencies $\Omega_{\pm}$ coincide.
The behaviours of the optical resonant frequencies 
under an increase of the power agree with the conclusion of the perturbative analysis deduced above. 
For $\phi=\pi/2 -0.47$, or more generally for $0<\phi<\pi/2$, 
the imaginary part of the optical resonant frequency  becomes
more negative when the laser power increases, and the resonance
becomes more stable; for  $\phi=\pi/2 -0.47$, 
or generically for $\pi/2<\phi<\pi$, the imaginary part becomes slightly 
less negative when the laser power increases.  
The behavior of the mechanical resonance is particularly interesting.
For $\phi=\pi/2-0.47$, or generically for $0<\phi<\pi/2$, and for very 
low laser power $I_0$ the two resonant frequencies separate along 
the real axis, as anticipated by the perturbative analysis. Moreover, 
as $I_0$ increases they both gain a positive 
imaginary part. However, since the trajectory is tangent to the real axis, 
the growth of the imaginary parts is much smaller than the growth of the real parts. 
For $\phi=\pi/2+0.47$, or more generally for $\pi/2<\phi<\pi$, the
two resonant frequencies separate along the imaginary axis, moving in that
direction as $I_0$ increases. 

We finally note that whenever the SR detuning
$\phi$ is different from $0$ and $\pi/2$, the mechanical 
resonance is always unstable. We shall discuss this issue in more detail 
in the next section.

\subsection{Characterization of mechanical instabilities}
\label{sec3.3}

As discussed in the previous section, the coupled mechanical 
resonant frequencies always have  a positive 
imaginary part, corresponding to an instability. 
The growth rate of this unstable mode is proportional to
the positive imaginary part of the resonant frequency. 
The time constant, or e-folding time of the mode, is $1/\Im(\Omega)$.
Hence, the larger the $\Im(\Omega)$ the more unstable the system is.

\begin{figure}[ht]
\begin{center}
\begin{tabular}{cc}
\hspace{-0.7cm}
\epsfig{file=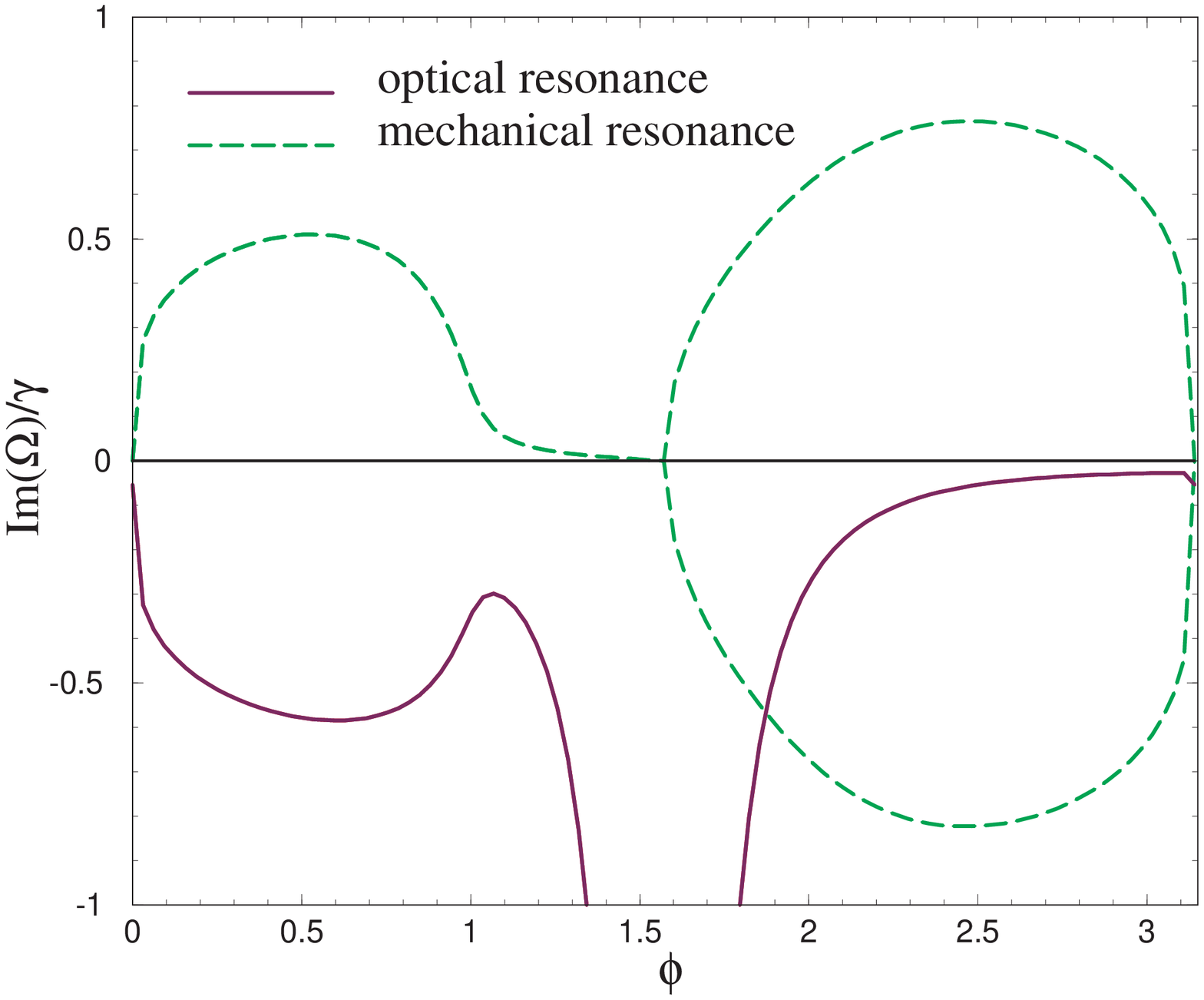,width=0.45\textwidth} &
\hspace{0.8cm}
\epsfig{file=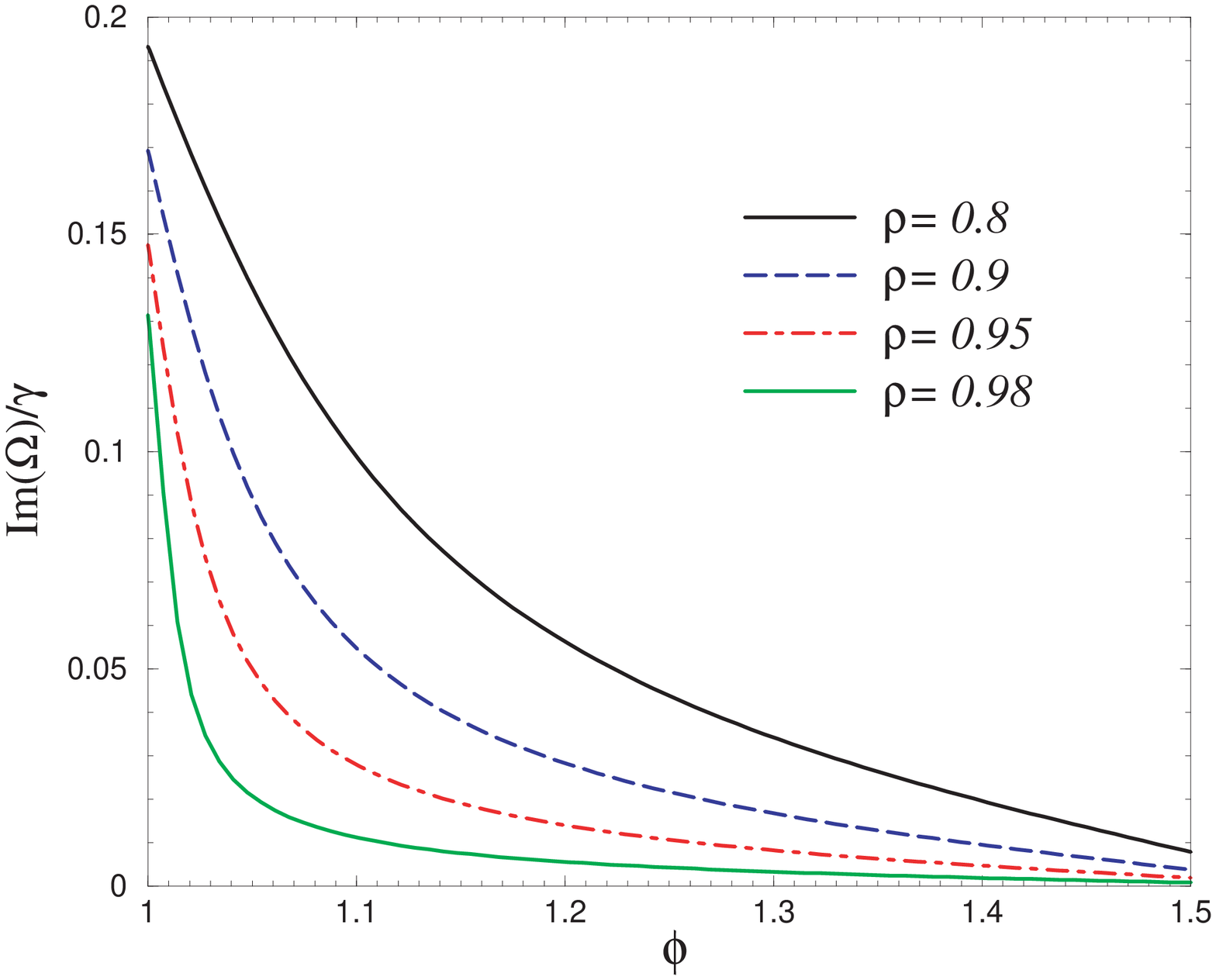,width=0.45\textwidth}
\end{tabular}
\caption{\sl The growth of instabilities for highly reflecting SR mirrors. 
In the left panel we plot the imaginary part of the
resonant frequencies, obtained solving Eq.~(\protect\ref{3.11Mod}),
versus the SR detuning phase $\phi$, for $\rho = 0.9$ and $I_0 =
I_{\rm SQL} \simeq 10^4\,{\rm W}$.
On the right panel we blow up the plot shown in the left panel
for the detuning region
${\cal D}= \{\phi: \arctan[(4I_0/I_{\rm SQL})^{1/3}] <
\phi < \pi/2 \}$, fixing $\rho = 0.8,0.9,0.95,0.98$ and $I_0 =
I_{\rm SQL}\simeq 10^4\,{\rm W}$.
This range of physical parameters corresponds to
interesting LIGO-II noise curves \protect\cite{BC1,BC2}.}
\label{Fig3}
\end{center}
\end{figure}

In order to quantify the consequences of the instability, we have
solved numerically the condition of resonances, Eq.~(\ref{3.11Mod}).
In the left panel of Fig.~\ref{Fig3}, we plot the imaginary parts of 
the four resonant frequencies, in units of $\gamma = {T c}/{4L}$ 
(the bandwidth of the arm cavity, see Sec.~\ref{identify}), 
as a function of the detuning phase $0<\phi<\pi$ of the SR cavity, 
fixing $I_0=I_{\rm SQL} \simeq 10^4\,{\rm W}$ and $\rho=0.9$. For an interferometer 
with arm-cavity length $L=4\,\rm km$, and internal-mirror power reflectivity $T=0.033$, 
which is the value anticipated by the LIGO-II community \cite{GSSW99}, we get 
$\gamma= 619\,\rm s^{-1}$. Hence, the storage time of the arm cavity is 
$1/\gamma \simeq 1.6\,\rm ms$.

{}From the left panel of Fig.~\ref{Fig3} we infer that the imaginary 
parts of the two coupled optical resonant
frequencies (shown with a solid line) coincide  
over the entire range $0 < \phi < \pi$. 
The imaginary parts of the two coupled mechanical resonant frequencies (drawn 
by a long-dashed line) also coincide for $0<\phi<\pi/2$, but 
they have opposite imaginary parts for $\pi/2<\phi<\pi$ 
(see also Fig.~\ref{FigResonance} for two special choices of $\phi$).
{}From the various plots we conclude that the region characterized by the weakest 
instability is $\phi\, \laq \,\pi/2$. It is important 
to note that for these values of the detuning phase the noise curves 
of a SR interferometer have two distinct valleys that beat the
SQL (see Sec.~IV of \cite{BC2}).
\footnote{~In this paper we are only concerned with the quantum noise. 
Thermal noise also contributes significantly  
to the total interferometer noise; for the current baseline design 
it is estimated to be slightly above the SQL \protect{\cite{BGV00}}, 
but design modifications are being explored \cite{ISA} which would 
reduce it to about half the SQL in amplitude.} 
In Ref.~\cite{BC2} the authors pointed out that the positions of the valleys of
the noise curves coincide roughly with the real parts of the system's coupled
mechanical and optical resonant frequencies. 
By taking into account Fig.~\ref{FigRFF} and the dynamics of the system, 
discussed in Sec.~\ref{origin}, we can make the following remark.
The ``spring constant'' $K(\Omega)$ is real only for $\Omega \ll \Omega_{\pm}$. 
For larger $\Omega$'s, its imaginary part contributes to that of the 
resonant frequency, and thus to the instability. 
Therefore, the farther the coupled mechanical resonant frequency is from the 
decoupled optical resonant frequency ($\Omega_{\pm}$), the less unstable it is. 
However, the distance between the coupled mechanical resonant frequency
and the decoupled optical resonant frequency ($\Omega_{\pm}$) is directly related to
the distance between the coupled mechanical and coupled optical resonant
frequencies. Therefore, the more separate the two coupled resonances are, i.e.\ 
the farther apart the two valleys of the noise curve are, the more stable 
the mechanical resonance is. 

In Ref.~\cite{BC2}, by analyzing the case of 
very highly reflecting SR mirrors ($\rho\rightarrow 1$) 
the authors found interesting noise curves for  
the detuning range ${\cal D}= \{\phi: \arctan[(4I_0/I_{\rm SQL})^{1/3}] < \phi < \pi/2 \}$
[see Sec.~IV A and, in particular, Eq.~(4.4) of Ref.~\cite{BC2}]. 
In the right panel of Fig.~\ref{Fig3}, we blow up the left panel
around this region ${\cal D}$ and plot various curves obtained by 
varying the SR reflectivity $\rho  = 0.8, 0.9,0.95$ and $0.98$. 
We observe that, for this parameter set, the largest growth rate 
is $\sim \! 0.2 \gamma \sim \! 124\,{\rm s}^{-1}$, corresponding 
to an e-folding time of $8$ ms, which is five times larger than the arm-cavity storage 
time.

Finally, we notice that the kind of instability we have found
in SR interferometers has an origin similar to the dynamical
instability induced in a detuned Fabry-Perot cavity by
the radiation-pressure force acting on the mirrors~\cite{All,MMPDHV,R00}.

\section{Control systems for signal recycled interferometers}
\label{sec5}

In this section we discuss how to suppress the instabilities 
present in SR interferometers by a suitable servo system. 
Since the control system must sense the mirror motion inside 
the observation band and act on (usually damp) it, there is an 
issue to worry about: If the dynamics is changed 
by the control system, it is not clear {\it a priori}  whether the resonant dips 
(or at least the mechanical one which corresponds to the unstable resonance),  
which characterize the noise curves in the uncontrolled SR interferometer~\cite{BC1,BC2},
will survive. In the following we shall show the existence of control systems
that suppress the instability without altering the noise curves 
of uncontrolled interferometers, thereby relieving ourselves from the above worry.

\subsection{Generic feed-back control systems: changing the
dynamics without affecting the noise}
\label{subsec5.1}

\begin{figure}
\vskip -0.5truecm
\begin{center}
\epsfig{file=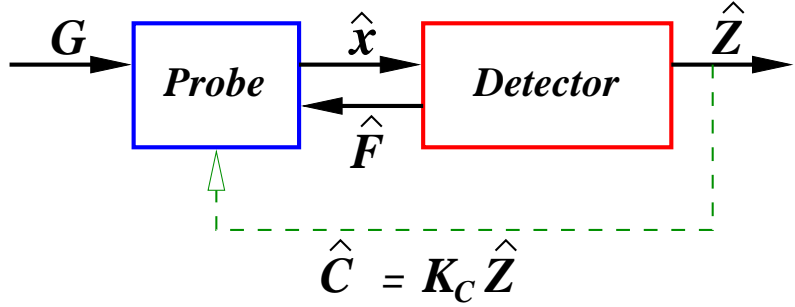,width=0.25\textwidth,angle=-90}
\vskip 0.2truecm
\caption{\sl Scheme of the control system introduced to 
quench the instabilities present in a SR interferometer. 
The output $\widehat{Z}$, which contains
the GW signal and the quantum noise, is sent through a linear filter 
with output $\widehat{\cal C} =
K_{\cal C}\,\widehat{Z}$, and is then
fed back onto the probe, i.e.\ the antisymmetric mode
of motion of the four arm-cavity mirrors.}
\label{Fig4}
\end{center}
\end{figure}

We shall identify a broad category of control systems for which, if
the instability can be suppressed, the noise curves are not altered.
We suppose that the output signal $\widehat{Z}$ is
sent through a linear filter $K_{\cal C}$ and then applied
to the antisymmetric mode of the arm-cavity mirrors
(see the schematic drawing in Fig.~\ref{Fig4}).
This operation corresponds to modifying the Hamiltonian (\ref{2.1}) into the form 
\beq
\widehat{H}= [(\widehat{H}_{\cal P}-\widehat{x}\,G)
+ \widehat{H}_{\cal D}] - \widehat{x}\,\widehat{F}
-\widehat{x}\,\widehat{\cal C}\,,
\label{hamc}
\eeq
where $\widehat{\cal C}$ is a detector observable whose free
Heisenberg operator (evolving under $H_{\cal D}$) at time $t$ 
is given, as required by causality, by an
integration over $t'< t$, 
\beq
\widehat{\cal C}^{(0)}(t) = \int_{-\infty}^t dt'\,K_{\cal
C}(t-t')\,\widehat{Z}^{(0)}(t')\,.
\label{3.13}
\eeq
Physically the filter kernel $K_{\cal C}(\tau)$ should be a function
defined for $\tau > 0$ and should decay to zero when $\tau \rightarrow
+\infty$. However, in order to apply Fourier analysis,
we can extend its definition to $\tau < 0$ by imposing $K_{\cal C}(\tau<0)\equiv
0$, thereby obtaining 
\beq
\widehat{\cal C}^{(0)}(t) = \int_{-\infty}^{+\infty} dt'\,K_{\cal
C}(t-t')\,\widehat{Z}^{(0)}(t').
\eeq
Therefore, in the Fourier domain we have
\beq
\widehat{\cal C}^{(0)}(\Omega)=K_{\cal C}(\Omega)
\widehat{Z}^{(0)}(\Omega),
\eeq
where $K_{\cal C}(\Omega)$ is the Fourier transform of $K_{\cal C}(\tau)$. 
It is straightforward to show that the two time-domain properties  
$K_{\cal C}(\tau<0)=0$ and $K_{\cal C}(\tau\rightarrow +\infty)\rightarrow 0$ 
correspond in the Fourier domain to the requirement that $K_{\cal
C}(\Omega)$ have poles only in the lower-half $\Omega$ plane.

Working in the Fourier domain and assuming that the readout scheme
is homodyne detection with detection phase $\zeta = {\rm const}$,  
we derive a set of equations of motion similar to Eqs.~(\ref{3.9})--(\ref{3.8}), 
\bea
\widehat{Z}_{\zeta}^{(1)}(\Omega)&=&
\widehat{Z}_{\zeta}^{(0)}(\Omega)+
\left[R_{Z_{\zeta}F}(\Omega) + 
R_{Z_{\zeta}{\cal C}_{\zeta}}(\Omega)\right]\,\widehat{x}^{(1)}(\Omega)
\label{2.8Cx}\,,\\
\widehat{F}^{(1)}(\Omega)&=& \widehat{F}^{(0)}(\Omega)+
\left[R_{FF}(\Omega)+R_{FC_{\zeta}}(\Omega)\right]\,\widehat{x}^{(1)}(\Omega)
\label{3.7Cx} \,,\\
\widehat{x}^{(1)}(\Omega)&=& R_{xx}(\Omega)\,
[G(\Omega)+\widehat{F}^{(1)}(\Omega) + \widehat{\cal C}_{\zeta}^{(1)}(\Omega)]
\label{3.8Cx} \,,\\
\widehat{\cal C}_{\zeta}^{(1)}(\Omega)&=&
\widehat{\cal C}_{\zeta}^{(0)}(\Omega)+
\left[R_{{\cal C}_{\zeta}F}(\Omega)+
R_{{\cal C}_{\zeta}{\cal C}_{\zeta}}(\Omega)\right]
\,\widehat{x}^{(1)}(\Omega) 
\label{evolutionCx} \,.
\eea
Each of Eqs.~(\ref{2.8Cx}), (\ref{3.7Cx}) and
(\ref{evolutionCx})  has two response
terms due to the two coupling terms between the probe and the
detector in the total 
Hamiltonian (\ref{hamc}). However, some of the responses are actually zero. 
In particular, inserting Eq.~(\ref{3.13}) into 
$[\widehat{F}^{(0)}(t),\widehat{\cal C}_\zeta^{(0)}(t')]$ and using the fact that 
$[\widehat{F}^{(0)}(t),\widehat{Z}_\zeta^{(0)}(t')]=0$ for $t>t'$  
[see Eq.~(\ref{2.14})], we find
$R_{FC_{\zeta}}(\Omega)=0$. Combining Eq.~(\ref{3.13}) with 
the fact that $[\widehat{Z}_\zeta^{(0)}(t),\widehat{Z}_\zeta^{(0)}(t')]=0$ for all $t,\,t'$
 [see Eq.~(\ref{2.14})], we have $R_{Z_{\zeta}{\cal C}_{\zeta}}(\Omega)=0=R_{{\cal C}_{\zeta}{\cal
C}_{\zeta}}(\Omega)$.
Moreover, the fact that $K_{\cal C}(t-t')=0=C_{Z^{(0)}F^{(0)}}(t,t')$
for $t<t'$ gives the equality 
$R_{{\cal C}_{\zeta}F}(\Omega)=K_{\cal C}(\Omega)\,R_{Z_{\zeta}F}(\Omega)$.
Imposing these conditions, we deduce a simplified set of equations of
motion:
\bea
\widehat{Z}_{\zeta}^{(1)}(\Omega)&=& \widehat{Z}_{\zeta}^{(0)}(\Omega)+
R_{Z_{\zeta}F}(\Omega)\,\widehat{x}^{(1)}(\Omega)\label{2.8C}\,,   \\
\widehat{F}^{(1)}(\Omega)&=&\widehat{F}^{(0)}(\Omega)+R_{FF}(\Omega)
\,\widehat{x}^{(1)}(\Omega) \label{3.7C} \,,\\
\widehat{x}^{(1)}(\Omega)&=& R_{xx}(\Omega)\,
[G(\Omega)+\widehat{F}^{(1)}(\Omega) + \widehat{\cal C}_{\zeta}^{(1)}(\Omega)]
\label{3.8C} \,, \\
\widehat{\cal C}_{\zeta}^{(1)}(\Omega)&=&K_{\cal C}(\Omega)\,\widehat{Z}^{(1)}(\Omega)
\label{evolutionC} \,.
\eea
Solving Eqs.~(\ref{2.8C})--(\ref{evolutionC}), we obtain
\bea
\label{2.12C}
&&\widehat{x}^{(1)}(\Omega)=
\frac{R_{xx}}{1-R_{xx}\,\left(R_{FF}+R_{Z_{\zeta} F}\,K_{\cal C}\right)}
\left[G(\Omega)+\widehat{F}^{(0)}(\Omega)
+K_{\cal C}(\Omega)\,\widehat{Z}_{\zeta}^{(0)}(\Omega)\right] \,, \\
\label{2.13C}
&& \widehat{Z}_{\zeta}^{(1)}(\Omega)=
\frac{1-R_{xx}\,R_{FF}}{1-R_{xx}\,\left(R_{FF}+R_{Z_{\zeta} F}\,
K_{\cal C}\right)}
\left\{\widehat{Z}_{\zeta}^{(0)}(\Omega)+
\frac{R_{Z_{\zeta}F}\, R_{xx}}{1-R_{xx}\,R_{FF}}\,
\left[G(\Omega)+\widehat{F}^{(0)}(\Omega)\right] \right\}\,,\\
&& \widehat{F}^{(1)}(\Omega)=
\frac{1-K_{\cal C}\,R_{xx}\,R_{Z_{\zeta}F}}{1-R_{xx}\,\left(R_{FF}+R_{Z_{\zeta} F}\,
K_{\cal C}\right)}
\left\{\widehat{F}^{(0)}(\Omega)+
\frac{R_{FF}\, R_{xx}}{{1-K_{\cal C}\,R_{xx}\,R_{Z_{\zeta}F}}}
\,\left[G(\Omega)+K_{\cal C}\,\widehat{Z}^{(0)}(\Omega)\right] \right \}\,.
\label{2.14C}
\eea
{}From the above equations (\ref{2.12C})--(\ref{2.14C}), we infer 
that the stability condition for the controlled system is determined 
by the positions of the roots of 
$[1-R_{xx}(R_{FF}+R_{Z_{\zeta}F}K_{\cal C})]$. Therefore,
by choosing the filter kernel $K_{\cal C}$ appropriately, it may be
possible that all the roots have negative imaginary part, in which case 
the system will be stable. 

Before working out a specific control kernel $K_{\cal C}$ that 
suppresses the instability, let us notice that different choices 
of $K_{\cal C}$ give outputs (\ref{2.13C}) that differ only by an overall 
frequency-dependent normalization factor. This factor 
does not influence the interferometer's
noise, since from Eq.~(\ref{2.13C}) we can see 
that the relative magnitudes of the signal (term proportional 
to $G$) and the noise (terms proportional to $\widehat{Z}^{(0)}_\zeta$ and 
$\widehat{F}^{(0)}$) depend only on the
quantities inside the brackets $\{\quad \}$ and
not on the factor multiplying the bracket [see Ref.~\cite{BC2} for a 
detailed discussion of the noise spectral density].
Therefore if this control system can suppress the instability,
the resulting well-behaved controlled SR interferometer 
will have the same noise as evaluated in Refs.~\cite{BC1,BC2} 
for the uncontrolled SR interferometer.
This important fact can be easily understood by observing
that, because the whole output (the GW signal $h$ and the noise $N$) is
fed back onto the arm-cavity mirrors, $h$ and $N$ are suppressed 
in the same way by the control system, and
thus their relative magnitude at any frequency $\Omega$ is 
the same as if the SR interferometer had been uncontrolled.  

\subsection{An example of a servo system: effective damping of the test-mass}
\label{subsec5.2}

Physically, it is quite intuitive to think of the feed-back system 
as a system that effectively ``damps'' the test-mass motion. 
When the control system is present, the equation of motion 
for the antisymmetric mode can be obtained from Eqs.~(\ref{3.8C}), 
(\ref{2.8C}) and (\ref{evolutionC}). It reads [as compared to Eq.~(\ref{3.8})]:
\beq
\widehat{x}^{(1)}(\Omega)= 
\frac{R_{xx}}{1 -K_{\cal C}\,R_{xx}\,R_{Z_{\zeta} F}}
\left[
G(\Omega) +
\widehat{F}^{(1)}(\Omega) + K_{\cal C}\,\widehat{Z}_{\zeta}^{(0)}(\Omega)
\right]
\label{effectiveX} \,.
\eeq
Denoting by ${R}^{\cal C}_{xx}$ the response of $\widehat{x}^{(1)}$ 
to $G$ and $\widehat{F}^{(1)}$ when the servo system is present, i.e.\   
\beq
{R}^{\cal C}_{xx} = \frac{R_{xx}}{1 -K_{\cal C}\,R_{xx}\,R_{Z_{\zeta} F}}\,,
\label{3.16}
\eeq
we can rewrite the overall normalization factor which appears in
Eqs.~(\ref{2.12C})--(\ref{2.14C}) as
\beq
\label{stabilityC}
\frac{1}{1-R_{xx}\,\left(R_{FF}+R_{Z_{\zeta} F}\,K_{\cal C}\right)}=
\frac{R^{\cal C}_{xx}}{R_{xx}}\,\frac{1}{1-R^{\cal C}_{xx} \,R_{FF}}\,.
\eeq
A sufficient condition for stability is that both ${R^{\cal C}_{xx}}/{R_{xx}}$
and ${1}/({1-R^{\cal C}_{xx} R_{FF}})$ have poles only in the lower-half
complex plane. [Note that when the servo system is present 
$R^{\cal C}_{xx}$ replaces $R_{xx}$ in the stability condition 
of the system, see Sec.~\ref{subsec2.2}, 
Eqs.~(\ref{2.11})--(\ref{2.13}) and discussions after them.]  

\begin{figure}[ht]
\vskip -0.4truecm
\begin{center}
\epsfig{file=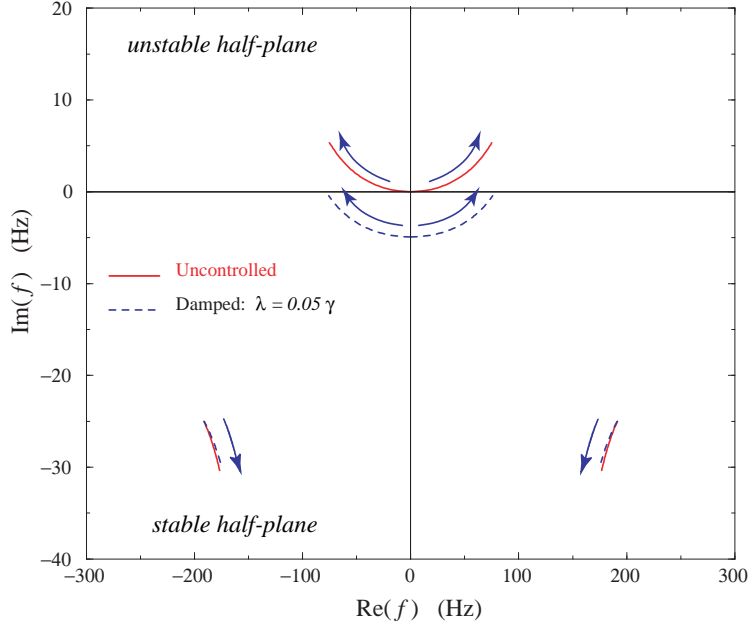,width=0.6\textwidth,angle=0}
\vskip 0.2truecm
\caption{\sl Effective damping due to a servo system 
with control kernel given by Eq.~(\protect\ref{3.19}).  
We have fixed: $\lambda=0.05\,\gamma$, $\rho=0.9$, 
$\phi=\pi/2-0.47$  and $I_0$ from $\sim
0$ up to $I_{\rm SQL}\simeq 10^4\,{\rm W} $. The arrows indicate the directions of
increasing light power $I_0$. 
The originally unstable mechanical resonance (solid line) 
is pushed downward in the complex $\Omega$-plane, 
and stabilized (dashed line). 
The figure also shows the effect of the control system on the stable optical resonances.} 
\label{FigResControl}
\end{center}
\end{figure}

We have found it natural to choose for $R^{\cal C}_{xx}(\Omega)$ the  
susceptibility of a damped oscillator (with effective mass $m/4$),
having both poles in the lower-half $\Omega$ plane at $\Omega=-i\lambda$, i.e.\
\footnote{~In the time domain this choice of $R^{\cal C}_{xx}(\Omega)$ 
corresponds to the equation of motion 
\beq
\frac{m}{4}\ddot{x}=- \frac{m \lambda }{2} \dot{x} - \frac{m \lambda^2
}{4} x + {\rm forces}\,.
\eeq
}
\beq
{R}^{\cal C}_{xx}(\Omega) = - \frac{4}{m}\,\frac{1}{(\Omega + i\lambda)^2}\,,
\label{3.18}
\eeq
with $\lambda$ a real parameter. This choice automatically ensures that
${R^{\cal C}_{xx}}/{R_{xx}}$ has poles only in the lower-half complex plane.
Moreover, by choosing $\lambda$ appropriately we can effectively push the roots
of $(1 -{R}^{\cal C}_{xx}\,R_{FF})$ in Eq.~(\ref{stabilityC}) to the lower-half $\Omega$
plane, as shown in Fig.~\ref{FigResControl} for $\rho =0.9$, $\phi = \pi/2-0.47$, 
$\lambda = 0.05\,\gamma$ and $I_0$ from
$\sim 0$ up to $I_{\rm SQL}$.

However, we also need to check that $K_{\cal C}(\Omega)$ has poles only 
in the lower-half $\Omega$ plane. Using Eqs.~(\ref{3.16}), (\ref{3.18}) 
we obtain the following explicit expression for the kernel:
\bea
\label{3.19}
K_{\cal C}(\Omega)
&=& 
\frac{1}{R_{Z_{\zeta}F}}
\left(
\frac{1}{R_{xx}}-
\frac{1}{R^{\cal C}_{xx}}
\right)
\nonumber 
\\
&=&\frac{m \lambda }{2 \, \tau }\sqrt{\frac{\hbar L^2}{2 I_0 \omega_0}}
\left(
\Omega+ \frac{i \lambda}{2}\right)
\frac{(1+2\rho\cos 2\phi +\rho^2)(\Omega-\Omega_-)(\Omega-\Omega_+)}
{(\Omega+i\gamma)\cos(\phi+\zeta)+\rho(\Omega-i\gamma)\cos(\phi-\zeta)}\,.
\eea
For $\zeta=0$ or $\zeta=\pi/2$, i.e.\
when either of the two quadratures $\widehat{b}_1$ or 
$\widehat{b}_2$ is measured,
the control kernel (\ref{3.19}) indeed has poles only in the lower-half 
complex plane. More generally, we have shown that if $0 < \phi < \pi/2$,
the control kernel (\ref{3.19}) has poles in the lower-half 
complex plane for all $\pi/2\le\zeta\le\pi$,
regardless of the value of $\rho$, but it may become unphysical in
the region $0<\zeta<\pi/2$.
However, for the unphysical values of $\zeta$ there are various
feasible ways out. For example, we could change $R^{\cal C}_{xx}$ 
by replacing $m$ in Eq.~(\ref{3.18}) with a slightly 
smaller quantity $m_{\cal C}$. In this case 
\beq
\left(\frac{1}{R_{xx}}-\frac{1}{R^{\cal C}_{xx}}\right) =
-\frac{m}{4}\,\left [ \Omega\,\left (1 -\sqrt{\frac{m_{\cal C}}{m}} \right )
- i\lambda\,\sqrt{\frac{m_{\cal C}}{m}} \,\right ]\,
\left [ \Omega\,\left (1 +\sqrt{\frac{m_{\cal C}}{m}} \right )
+ i\lambda\,\sqrt{\frac{m_{\cal C}}{m}} \,\right ]\,.
\label{m}
\eeq
By choosing $m_{\cal C}$ appropriately, we can use the first factor
in Eq.~(\ref{m}), which has a root in the upper-half complex
plane, to cancel the bad pole coming from $R_{Z_{\zeta}F}$ in Eq.~(\ref{3.19}), 
so that $K_{\cal C}$ will have poles only in the lower-half 
complex plane. Finally, we must  adjust $\lambda$ 
so that the effective damping suppress the instability.

Of course, the servo electronics employed 
to implement the control system will inevitably introduce 
some noise into the interferometer. In our investigation 
we have not modelled this noise. However, LIGO experimentalists 
have seen no fundamental noise limit 
in implementing control kernels of the kind we discussed, 
and deem it technically possible to suppress any 
contribution coming from the electronics to within 10\% of the total 
predicted quantum noise \cite{Ken,Nergis}.
This issue deserves a more careful study and it will be tackled 
elsewhere \cite{BCM}.

In this paper we have restricted ourselves to the readout scheme 
of frequency independent homodyne detection, 
in which only one (frequency independent) quadrature $b_{\zeta}$ is measured. 
The issue of control-system design when other readout schemes 
are present, e.g., the so-called radio-frequency 
modulation-demodulation design, is currently under investigation \cite{BCM}.

Finally, for simplicity we have limited our discussion to lossless SR
interferometers.  When optical losses are taken into account, 
we have found that the instability problem is still
present \cite{BC2} and we have checked that those instabilities
can be cured by the same type of control system as was discussed
above for lossless SR interferometers. 

\section{Conclusions}
\label{sec6}

Using the formalism of linear quantum-measurement
theory, extended by Braginsky and
Khalili \cite{BK92} to GW detectors, we have described the optical-mechanical 
dynamics of SR interferometers such as LIGO-II \cite{GSSW99}. 
This analysis has allowed us to work out various significant 
features of such interferometers, which previous investigations 
\cite{D82,VMMB88,M88,MSNCSRWD93,M95} could not reveal.

We have found that when the (carrier) laser frequency is 
detuned in the SR cavity, the arm-cavity mirrors are 
not only perturbed by a random fluctuating force
but are also subject to a linear restoring force with a
specific frequency-dependent rigidity.
This phenomenon is not unique to SR interferometers; 
it is a generic feature of detuned cavities \cite{All,OB,MMPDHV,R00} and 
was originally used by Braginsky, Khalili and colleagues in 
designing the ``optical bar'' GW detectors~\cite{OB}.

Our analysis has revealed that, for SR interferometers, 
the dynamics of the whole optical-mechanical system, composed of 
the arm-cavity mirrors and the optical field, resembles 
that of a free test mass (mirror motion) connected 
to a massive spring (optical fields). 
When the test mass and the spring are not connected (e.g., for very low laser power) 
they have their own eigenmodes, namely the uniform
translation mode for the free test mass (free antisymmetric mode), and the 
longitudinal-wave
mode for the spring (decoupled SR optical resonance). However, as soon as the free 
test mass is connected to the massive spring (e.g, for LIGO-II laser
power), the two free modes 
get shifted in frequency, so the entire coupled
system can resonate at two pairs of finite frequencies (coupled mechanical and
optical resonances). 
{}From this point of view a SR interferometer behaves 
like an ``optical spring'' detector. For LIGO-II parameters,
both resonant frequencies can lie in the observation band 
$10\,{\rm Hz} < f < 10\,{\rm kHz}$ and they are 
responsible for the beating of the SQL in SR 
interferometers ~\cite{BC1,BC2}.

The formalism used in the present paper has allowed 
us to analyze in more detail the features of the instabilities 
in SR interferometers, pointed out in Refs.~\cite{BC1,BC2}. 
Most importantly, we have shown the possibility of using a
feed-back control system to cure such instabilities without
compromising the performance of the interferometer. 
However, before any practical implementation, 
a much more careful and precise study should be carried out,  
including various readout schemes \cite{BCM}.

Finally, the general discussion based on the Braginsky-Khalili 
force-susceptibility formalism, given in the first part 
of this paper (Sec.~\ref{sec2}), and the application to 
a specific type of GW interferometer, the LIGO-II SR interferometer, 
given in the second
part of it (Secs.~\ref{sec3}--\ref{sec5}), may provide, along with
Refs.~\cite{KLMTV00,BC2}, a framework for future 
investigations of quantum noise in advanced, more complex,  
optical configurations.

\acknowledgments
We wish to thank P.~Fritschel, J.~Mason, N.~Mavalvala, G.~Mueller and K.A.~Strain 
for very interesting, helpful discussions and/or comments.
It is also a pleasure to thank V.B.~Braginsky for pointing
out the importance of optical-mechanical oscillations
in GW detectors, F.Ya.~Khalili
for very stimulating interactions concerning the
optical-mechanical rigidity in LIGO-II and Yu.~Levin for very
lively discussions which further motivated our descriptions
 of SR interferometers using the force-susceptibility
approach. Finally, we are deeply indebted to K.S.~Thorne for 
his constant support and for offering  numerous useful 
comments and suggestions.

This research was supported by NSF
grants PHY-9900776 and PHY-0099568 
and also for AB by Caltech's Richard Chase Tolman Fellowship.

\newpage
\appendix
\section{Basic properties of linear systems}
\label{appA}
In this Appendix, to clarify the formalism used in Sec.~\ref{sec2}, 
we summarize some well-known basic properties of linear systems linearly coupled
to each other or to external classical forces. Much of this material can be found in 
Sakurai \cite{sakurai},  and for its application to quantum-measurement processes
in Braginsky and Khalili \cite{BK92} and Caves et al.\ \cite{CTDSZ80}. 

\begin{definition}[Linear systems]
Any system whose Hamiltonian is at most quadratic in
its canonical coordinates and momenta is a linear system.
\end{definition}
\begin{definition}[Linear observables]
Any linear combination (either time dependent or time independent)
of the canonical coordinates and momenta of a linear system, plus a possible 
complex number (C-number), is a linear observable of the system.
\end{definition}

Denoting all the canonical coordinates and momenta by
$\widehat{{\cal C}}_i$ with $i=1,2,\cdots$, the Hamiltonian of a
linear system can be written as
\beq
\label{a1}
\widehat{H}(t)=\sum_{i,j}L^{ij}_{2}(t)\,\widehat{\cal C}_i\,\widehat{\cal C}_j+
\sum_{i}L^{i}_{1}(t)\,\widehat{{\cal C}}_i+L_0(t)\,,
\eeq
where $L^{ij}_2(t)$ is symmetric in $i$ and $j$.
The equations of motion of the canonical observables in the Heisenberg picture read 
[we use the fact that $\widehat{\cal C}_{jH}$ does not depend explicitly on time]:
\bea
\label{a2}
i \hbar \frac{d}{dt}\widehat{\cal C}_{jH}(t)
&=&\left[\widehat{\cal C}_{jH}(t),\widehat{H}_{H}(t)\right]\,, \nonumber \\
&=&\widehat{U}^{\dagger}(-\infty,t)\,
\left[\widehat{\cal C}_{j S},\widehat{H}_{S}(t)\right]\,\widehat{U}(-\infty,t)\,,  \nonumber \\
&=&\widehat{U}^{\dagger}(-\infty,t) \left[
\sum_{l,m}2\,L_2^{lm}(t)\,C_{jl}\,\widehat{\cal C}_{m S}
+\sum_{l}L_1^l(t)\,C_{jl} \right] \widehat{U}(-\infty,t)\,, \nonumber \\
&=&\sum_{l,m}2\,L_2^{lm}(t)\,C_{jl}\,\widehat{\cal C}_{m H}(t)
+\sum_{l}L_1^l(t)\,d_{jl}\,.
\eea
Here
the subscripts $S$ and $H$ stand for Schr\"odinger and Heisenberg pictures
respectively, $C_{jl}\equiv [\widehat{\cal C}_{j S},\widehat{\cal C}_{l S}]$
is the commutator
between the canonical operators, which is a C-number, and 
$\widehat{U}(-\infty,t)$ is the time-evolution operator which satisfies the 
Schr\"odinger equation
\beq
\label{a3}
i \hbar \frac{d}{dt}\widehat{U}(-\infty,t)
=\widehat{H}_{S} \, \widehat{U}(-\infty,t)
\eeq
with initial condition $\widehat{U}(-\infty,-\infty)=1.$
The solution to Eq.~(\ref{a2}) is of the form
\beq
\label{solution}
\widehat{\cal C}_{j H}(t)=\sum_{k}\alpha_{jk}(t)\,
\widehat{\cal C}_{k H}(-\infty)+\beta_{j}(t)
=\sum_{k}\alpha_{jk}(t)\,\widehat{\cal C}_{k S}+\beta_{j}(t)\,,
\eeq
where
$\alpha_{jk}(t)$ and $\beta_{j}(t)$ are time dependent C-numbers.

For any linear observable $A$ it follows from linearity  that
$\widehat{A}_H(t) = \sum_j a_j(t)\, \widehat{\cal C}_{j H}(t)+b(t)$,
which, along with Eq.~(\ref{solution}), leads to:
\beq
\label{a6}
\widehat{A}_{H}(t)=\sum_j a_j(t)\,\widehat{\cal C}_{jH}(t)+b(t)=
\sum_{j,k} a_j(t)\, \alpha_{jk}(t)\, \widehat{\cal C}_{k S}
+\sum_j a_j(t) \beta_j(t)+b(t)
\,.
\eeq
This provides the following theorem:

\begin{theorem}
At any time the operator of a linear
observable in the Heisenberg picture can always be
written as a linear combination of operators of the
(time-independent) canonical variables
in the Schr\"odinger picture plus a possible C-number.
\end{theorem}

Applying the above theorem to any two linear observables $A$ and $B$, recalling that
$C_{jk}\equiv [\widehat{\cal C}_{j S},\widehat{\cal C}_{k S}]$
is a C-number and the commutator between a C-number and any operator is zero, we find
\beq
\label{a7}
\left[\widehat{A}_{H}(t),\widehat{B}_{H}(t')\right]=
\sum_{j,k} \gamma_j^A(t)\, \gamma^B_k(t')\, C_{jk}\,,
\eeq
which is a C-number. Therefore, the following theorem holds:
\begin{theorem}
In the Heisenberg picture, the commutator of the operators of any
two linear observables at two times is a C-number.
\end{theorem}

We are interested in the evolution of a linear system subject to a classical
external linear force or
linearly coupled to another independent linear system. A force-susceptibility kind
of formulation can be introduced in these cases (as is done by Braginsky and
Khalili, see Sec.~6.4 of Ref.~\cite{BK92}).
We shall describe the system using a perturbative approach. Thus we write the total
Hamiltonian in the Schr\"odinger picture as $\widehat{H}_S=\widehat{H}_{0S}+\widehat{V}_S(t)$,
where $\widehat{V}_S(t)$ is treated as a perturbation with respect
to the zeroth order Hamiltonian $\widehat{H}_{0 S}$.
It is generally convenient to introduce the so-called Interaction
picture (see, e.g., Sections 5.5 and 5.6 of Ref.~\cite{sakurai}), 
in which the evolution operator $\widehat{U}_{I}$ is 
defined by the relation $\widehat{U}(-\infty,t)\equiv\widehat{U}_0(-\infty,t)\,\widehat{U}_{I}(-\infty,t)
$, where $\widehat{U}_0(-\infty,t)$ is the evolution operator associated
with $\widehat{H}_{0 S}$ and $\widehat{U}$ is defined by Eq.~(\ref{a3}).
Then, $\widehat{U}_{I}(-\infty,t)$ satisfies the equations
\beq
\label{a11}
i \hbar \frac{d}{dt}\widehat{U}_{I}(-\infty,t)=
\widehat{V}_{I}(t)\,\widehat{U}_{I}(-\infty,t)\,, \quad \quad
\widehat{U}_{I}(-\infty,-\infty)=1\,,
\eeq
with $\widehat{V}_{I}(t)\equiv
\widehat{U}^{\dagger}_0(-\infty,t)\,\widehat{V}_{S}(t)\,
\widehat{U}_0(-\infty,t)$.
The solution of Eq.~(\ref{a11}) can be written as a
perturbative expansion,
\bea
\widehat{U}_{I}(-\infty,t) &=&
1+\frac{1}{i \hbar}\int_{-\infty}^t dt_1 \widehat{V}_{I}(t_1)
 +\left(\frac{1}{i \hbar}\right)^2
\int_{-\infty}^t dt_1 \int_{-\infty}^{t_1} dt_2 \widehat{V}_{I}(t_1) \widehat{V}_{I}(t_2)
 + \cdots\,, \nonumber \\
&=&
\sum_{n=0}^{\infty}\frac{1}{n!}\left(\frac{1}{i \hbar}\right)^n
T\left\{\left[\int_{-\infty}^t dt_1  \widehat{V}_{I}(t_1)\right]^n\right\}\,,
\label{a13}
\eea
where $T$ denotes the time-ordered product \cite{merzbacher}.
The Heisenberg operator associated with
any observable $A$, evolving under the full Hamiltonian $\widehat{H}$,
is linked to the corresponding Heisenberg operator
evolving under the Hamiltonian $\widehat{H}_0$ by the relation
$ \widehat{A}_{H}(t)=
\widehat{U}_{I}^{\dagger}(-\infty,t)\widehat{A}_{H}^{(0)}(t)
\widehat{U}_{I}(-\infty,t)$,
where the superscript $(0)$ on the observable $A$ denotes that the evolution is due to
$\widehat{H}_0$. Inserting Eq.~(\ref{a13}) into the above equation,
we get
\bea
&& \widehat{A}_{H}(t)=\widehat{A}_{H}^{(0)}(t) +
\frac{i}{\hbar}\int_{-\infty}^t dt_1 \left[\widehat{V}_{I}(t_1),\widehat{A}^{(0)}_{H}(t)\right]+
\left(\frac{i}{\hbar}\right)^2 \int_{-\infty}^t dt_1 \int_{-\infty}^{t_1} dt_2
\left[\widehat{V}_{I}(t_2),\left[\widehat{V}_{I}(t_1),\widehat{A}^{(0)}_{H}(t)\right]\right]
\nonumber \\
&& + \cdots +\left(\frac{i}{\hbar}\right)^n
\int_{-\infty}^t dt_1 \int_{-\infty}^{t_1} dt_2 \cdots \int_{-\infty}^{t_{n-1}} dt_n
\left[\widehat{V}_{I}(t_n),\left[\cdots,\left[\widehat{V}_{I}(t_2),\left[\widehat{V}_{I}(t_1),
\widehat{A}^{(0)}_{H}(t)\right]\right]\cdots\right]\right] + \cdots \,.
\label{a15}
\eea
For a linear system subject to an external classical linear force $G(t)$,
the interaction term is $\widehat{V}_{I}(t)=-\widehat{x}_{H}^{(0)}G(t)$.
Plugging this expression into Eq.~(\ref{a15}) and using Theorem 2, it is straightforward
to deduce that the second and all higher order terms in Eq.~(\ref{a15})
vanish and the first order perturbation gives the exact solution.
Hence, we obtain the following theorem:
\begin{theorem}
Consider a linear system subject to a classical generalized force $G(t)$, whose
Hamiltonian is given by $\widehat{H}=\widehat{H}_0-\widehat{x}\,G(t)$,
where $\widehat{x}$ is a linear observable. Then, for any linear observable $\widehat{A}$,
the Heisenberg operator $\widehat{A}_{H}(t)$
can be written as
the sum of its free-evolution part, $\widehat{A}^{(0)}_{H}(t)$, plus a term which is due
to the presence of the external force, i.e.\
\beq
\label{a17}
\widehat{A}_{H}(t)=
{\widehat{A}^{(0)}}_{H}(t)+\frac{i}{\hbar}\int^t_{-\infty}dt'\,C_{Ax}(t,t')\,G(t')\,,
\eeq
where $C_{Ax}(t,t')$ is a C-number, called the (time-domain) susceptibility,
given explicitly by
\beq
\label{a18}
C_{Ax}(t,t')\equiv[\widehat{A}_{H}^{(0)}(t),\widehat{x}_{H}^{(0)}(t')]\,.
\eeq
\end{theorem}

Let us now suppose that we have two independent linear systems
${\cal P}$ (e.g., the probe) and ${\cal D}$ (e.g., the detector), which by
definition are described by two different Hilbert spaces
$\cal{H}_{\cal P}$ and $\cal{H}_{\cal D}$.
We introduce the Hilbert space ${\cal H} = {\cal H}_{\cal P} \otimes {\cal H}_{\cal  D}$ and
define for  any operator $\widehat{x}$ of the system ${\cal P}$
the corresponding operator acting on ${\cal H}$ as $\widehat{x} \otimes \widehat{1}$,
while for any operator $\widehat{F}$ of the system
${\cal D}$ we introduce the operator $\widehat{1}\otimes\widehat{F}$ which
acts on ${\cal H}$.
Henceforth, we shall limit ourselves to interaction terms $V$,
in the total Hamiltonian $\widehat{H}=\widehat{H}_{\cal P}+
\widehat{H}_{\cal D}+\widehat{V}$,
of the form: $\widehat{V}=-\widehat{x}\otimes\widehat{F}$,
with $\widehat{x}$ and $\widehat{F}$ acting on ${\cal P}$ and ${\cal D}$, respectively.
Using Eq.~(\ref{a15}) with
$\widehat{V}_{I}(t)=-\widehat{x}_{H}^{(0)}(t) \widehat{F}_{H}^{(0)}(t)$,
noticing that (i) the zeroth order Heisenberg operators
of two observables living in different Hilbert spaces
commute and (ii) the zeroth order Heisenberg operators of two linear
observables living in the same Hilbert space have a C-number commutator,
we derive the following theorem:
\begin{theorem}
Consider two independent linear systems ${\cal P}$ and ${\cal D}$, and two
linear observables, $\widehat{x}$ of ${\cal P}$ and $\widehat{F}$ of ${\cal D}$.
Suppose that the two systems are coupled by a term
$-\widehat{x}\otimes\widehat{F}$, i.e.\ the Hamiltonian of the composite system ${\cal P}$
+ ${\cal D}$ reads $\widehat{H}=\widehat{H}_{\cal P}+
\widehat{H}_{\cal D}-\widehat{x}\otimes\widehat{F}.$
Then, for any linear observable $\widehat{A}$ of the system ${\cal P}$
and $\widehat{B}$ of the system ${\cal D}$, their full Heisenberg evolutions are given by:
\beq
\widehat{A}_{H}(t)=\widehat{A}^{(0)}_{H}(t)+\frac{i}{\hbar}\,
\int^t_{-\infty}dt'C_{Ax}(t,t')\,\widehat{F}_{H}(t')\,, \quad \quad
\widehat{B}_{H}(t)=\widehat{B}^{(0)}_{H}(t)+\frac{i}{\hbar}\,
\int^t_{-\infty}dt'C_{BF}(t,t')\,\widehat{x}_{H}(t')\,,
\label{a20}
\eeq
where $\widehat{A}^{(0)}_{H}$ and $\widehat{B}^{(0)}_{H}$ stand for
the free Heisenberg evolutions, and the susceptibilities are defined by
\beq
C_{Ax}(t,t') \equiv [\widehat{A}_{H}^{(0)}(t),\widehat{x}_{H}^{(0)}(t')]\,,
\quad \quad
C_{BF}(t,t') \equiv [\widehat{B}_{H}^{(0)}(t),\widehat{F}_{H}^{(0)}(t')]\,.
\label{a21}
\eeq
\end{theorem}
In the case where the zeroth order Hamiltonian is time independent,
it is easy and convenient to express the above formalism in the Fourier domain.
We first notice that for a time independent $\widehat{H}_0$,
$\widehat{U}_0(t,t+\tau)=e^{-i \widehat{H}_0 \tau / \hbar }$
and for any two linear observables
$\widehat{A}_1$ and $\widehat{A}_2$ we have
$C_{A_1 A_2}(t+\tau,t'+\tau) = C_{A_1 A_2}(t,t')$,
i.e.\ $C_{A_1 A_2}(t,t')$ depends only on $t-t'$.
Defining the Fourier transform of any observable $\widehat{A}(t)$ as
\beq
\label{a23}
\widehat{A}(\Omega)\equiv\int_{-\infty}^{+\infty}dt \, e^{i\Omega t}\,\widehat{A}(t)\,,
\eeq
Eq.~(\ref{a17}) becomes $\widehat{A}_{H}(\Omega)=\widehat{A}^{(0)}_{H}(\Omega)+
R_{Ax}(\Omega)\,G(\Omega)$
while Eq.~(\ref{a20}) can be recast in the form
\beq
\label{a25}
\widehat{A}_{H}(\Omega)=\widehat{A}^{(0)}_{H}(\Omega)+R_{Ax}(\Omega)\,
\widehat{F}_{H}(\Omega)\,,
\quad \quad \widehat{B}_{H}(\Omega)=\widehat{B}^{(0)}_{H}(\Omega)+R_{BF}(\Omega)\,
\widehat{x}_{H}(\Omega)\,,
\eeq
where $R_{AB}(\Omega)$ is the susceptibility in the Fourier-domain, given by
\beq
\label{a26}
R_{AB}(\Omega)
=\frac{i}{\hbar}\int_{-\infty}^{+\infty}d\tau\,e^{i\Omega\tau}\,\Theta(\tau)C_{AB}(0,-\tau)
=\frac{i}{\hbar}\int_{0}^{+\infty}d\tau\,e^{i\Omega\tau}\,C_{AB}(0,-\tau)\,,
\eeq
with $\Theta(\tau)$ the step function.
For future reference, let us point out two properties which $R_{AB}(\Omega)$ satisfies and
that we use repeatedly in Sec.~\ref{sec2}:
\beq
\label{a27}
R^*_{AB}(\Omega)=R_{AB}(-\Omega)\,,
\quad \quad \left[\widehat{A}^{(0)}_{H}(\Omega_1),\widehat{B}^{(0)}_{H}(\Omega_2)\right]=
- 2 \pi i \hbar
\delta(\Omega_1+\Omega_2)\left[R_{AB}(\Omega_1)-R_{BA}(\Omega_2)\right]\,.
\eeq
To deduce the first identity in Eq.~(\ref{a27}), we consider 
the complex (Hermitian) conjugate of Eq.~(\ref{a26})
and use the Hermiticy of $\widehat{A}^{(0)}_H(t)$ and $\widehat{B}^{(0)}_H(t)$. 
For the second identity in Eq.~(\ref{a27}), we take the double Fourier transform of
$[\widehat{A}^{(0)}_H(t_1),\widehat{B}^{(0)}_H(t_2)]$ with respect to $t_1$ and $t_2$, 
and then using Eq.~(\ref{a26}) we find that the region corresponding to $t_1>t_2$ in 
the double integral yields the $R_{AB}$ term of Eq.~(\ref{a27}), 
while the region corresponding 
to $t_1<t_2$ gives the $R_{BA}$ term.

\end{document}